\documentclass[12pt]{article}

\usepackage{xspace,colortbl}
\usepackage{graphicx}
\usepackage{amsmath}
\usepackage{amsthm}
\usepackage{amssymb}
\usepackage{indentfirst}
\usepackage{setspace}
\usepackage[top=1.2in,bottom=1in]{geometry}
\usepackage{mathrsfs}

\begin{document}

\title{\bfseries \large{INSTABILITIES INDUCED BY PHASE FRONTS COALESCENCE DURING THE PHASE TRANSITIONS IN A THIN SMA LAYER: MECHANISM AND ANALYTICAL DESCRIPTIONS}}

\author{\small Hui-Hui Dai$^1$ and Jiong Wang$^2$
\\\small $^1$Department of Mathematics and Liu Bie Ju Centre for Mathematical Sciences,\\\small City University of Hong Kong,\ 83 Tat Chee Avenue, Kowloon Tong, Hong
Kong\\ \small Email: mahhdai@cityu.edu.hk
\\\small $^2$Department of
Mathematics, City University of Hong Kong, 83 Tat Chee
Avenue,\\\small Kowloon Tong, Hong Kong\\ \small Email:
jionwang@cityu.edu.hk }
\date{}
\maketitle


\begin{abstract}

Systematic experiments on stress-induced phase transitions in thin
SMA structures in literature have revealed two interesting
instability phenomena: the coalescence of two martensite-austenite
fronts leads to a sudden stress drop and that of two
austenite-martensite fronts leads to a sudden stress jump. In order
to get an insight into these two phenomena, in this work we carry
out an analytical study on the stress-induced phase transitions in a
thin SMA layer (a simple structure in which the two coalescence
processes can happen). We derive a quasi-2D model with a non-convex
effective strain energy function while taking into account the
rate-independent dissipation effect. By using a coupled
series-asymptotic expansion method, we manage to express the total
energy dissipation in terms of the leading-order term of the axial
strain. The equilibrium equations are obtained by maximizing the
total energy dissipation, which are then solved analytically under
suitable boundary conditions. The analytical results reveal that the
mechanism for such instabilities is the presence of "limited
points", which cause the switch of nontrivial solution modes to
trivial solution modes. Descriptions for the whole coalescence
processes of two fronts are also provided based on the analytical
solutions, which also capture the morphology varies of the specimen.
It is also revealed the key role played by the thickness-length
ratio on these instabilities: the zero limit of which can lead to
the smooth switch of nontrivial modes to trivial modes with no
stress drop or stress jump.

\end{abstract}



\section{Introduction}

Systematic experiments on stress-induced phase transitions in thin
SMA structures, such as wires, strips and tubes, have been carried
out (Lexcellent $\&$ Tobushi 1995; Shaw $\&$ Kyriakides 1995, 1997;
Sun \emph{et al} 2000; Favier \emph{et al} 2001; Li $\&$ Sun 2002;
Feng $\&$ Sun 2006). In general, the whole experimental process can
be divided into three stages. In the first stage, the deformation of
the specimen is the elastic deformation of initial phase. In the
second stage, the product phase first nucleates at some special
site, and then propagates gradually along the specimen. Finally, the
phase fronts coalescence and the whole length of the specimen has
transformed to the product phase. In the third stage, the
deformation of the specimen is the elastic deformation of the
product phase. It was also found that the measured engineering
stress-strain curves have some key features, e.g., the nucleation
stress peak (for the loading case) and stress valley (for the
unloading case), the stress plateaus, the rate-independent
hysteresis loop and so on.

In this paper, we shall focus on the instability phenomena induced
by the coalescence of two phase fronts. During the experiments,
these instability phenomena can be easily identified from both the
stress-strain curves and the surface morphology of the specimens
(Shaw $\&$ Kyriakides 1997; Feng $\&$ Sun 2006). Systematic
experimental results have obviously shown that the coalescence
process is inevitably accompanied the varies of the stress value and
the surface morphology of the specimen. In general, the coalescence
process has the following procedure as described in Feng $\&$ Sun
(2006). In the loading (unloading) case, as the total elongation
increases (decreases), two martensite-austenite
(austenite-martensite) phase fronts move towards each other. The
austenite (martensite) region sandwiched between the phase fronts
becomes more and more narrow. Eventually these two fronts start
getting touch each other and the coalescence process takes place. It
is clear the coalescence process is a dynamic process and seemingly
uncontrollable in the sense that it occurs very rapidly (this may
indicate that an instability occurs). During the coalescence
process, the local configuration of the specimen transforms from an
inhomogeneous mode to a homogeneous mode and the corresponding
stress value has a rapid drop for the loading case and a rapid jump
for the unloading case.

In order to get an insight into the phase fronts coalescence
phenomena, in this work we carry out an analytical study on the
stress-induced phase transitions in a thin SMA layer (a simple
structure in which the two coalescence processes can happen). We
shall derive a quasi-2D continuum model with a non-convex effective
strain energy function while taking into account the
rate-independent dissipation effect. The starting point of our model
is the formulation of Rajagopal $\&$ Srinivasa (1999, 2004) (also
see the formulation of Sun $\&$ Hwang 1993a, b). In fact, we have
already proposed a quasi-3D continuum model earlier to study the
phase transitions induced by extension in a slender SMA cylinder
(see Wang $\&$ Dai 2009), which has almost the same formulation with
this current model. Thus, we shall only give a brief introduction to
the derivation procedure in this paper and the detailed derivation
procedure can be found in Wang $\&$ Dai (2009).

In the experiments, typically the thickness-length (or width-length
or radius-length) ratio of the specimen is of order $O(10^{-2})$. As
a result, one might think that the deformation along the lateral
direction can be neglected and the layer can be treated as a
one-dimensional object. However, sometimes a purely one-dimensional
model appears to be not sophisticated enough to capture some key
features observed in experiments. Some explanations have already
been given in Dai $\&$ Cai (2006). In this paper, the total elastic
potential energy of the layer will be considered based on a
two-dimensional setting. Starting from the two-dimensional governing
system and by using the coupled series-asymptotic expansion method
(Dai $\&$ Cai 2006; Cai $\&$ Dai 2006), we manage to express the
total elastic potential energy of the layer in terms of the leading
order term of the axial strain (cf. (3.41)). Although the final
expression of the total elastic potential energy is one-dimensional,
it takes into account the lateral deformation and has the
higher-dimensional effects built in.

To describe the hysteretic behavior during the phase transition
process, one also needs to consider the inevitable mechanical
dissipation effect. In this paper, the mechanical dissipation effect
will be considered in a purely one-dimensional setting, i.e., we
neglect the influence of the radial deformation on the mechanical
dissipation. A specific constitutive form of the rate of dissipation
function is adopted in our model (cf. (2.6)). Based on the phase
transition criteria and the criterion of maximum rate of
dissipation, we propose the evolution laws of the phase state
variable for the purely loading and purely unloading processes.
After doing some further analysis, we derive the one-dimensional
expressions for the mechanical dissipation functions in terms of the
axial strain (cf. (4.5)-(4.8)).

With the expressions of the total elastic potential energy and the
mechanical dissipation functions, the equilibrium configurations of
the layer for the purely loading and purely unloading processes can
be determined by using the principle of maximizing the total energy
dissipation. By using the variational method, we derive the
equilibrium equations, which are then solved analytically under
suitable boundary conditions. It will be seen that the solutions
obtained show qualitatively agreements with the experimental
results.

Based on the analytical solutions obtained and by using the
limit-point instability criterion, we further consider the phase
fronts coalescence process. It is revealed that during the
coalescence process, the configurations of the layer switched from
the nontrivial solution modes to the trivial solution modes, which
is caused by the presence of the ``limit points". The morphology
varies of the layer and the accompanying stress drop/jump during the
coalescence process can be described. The influence of the
thickness-length ratio of the specimen on the coalescence process is
also studied. It will be shown that the zero limit of the
thickness-length ratio can lead to the smooth switch of nontrivial
solutions to trivial solutions with no stress drop or stress jump.

This paper is arranged as follows. In section 2, we give a simple
introduction to the formulation of Rajagopal $\&$ Srinivasa (1999,
2004) and propose the evolution laws of the phase state variable for
the purely loading and purely unloading processes. In section 3, we
formulate the field equations by treating the thin layer as a
two-dimensional object. By using the coupled series-asymptotic
expansion method, we express the total elastic potential energy of
the layer in terms of the asymptotic axial strain. In section 4, we
study the mechanical dissipation effect. After some analysis, we
express the mechanical dissipation functions in terms of the axial
strain. In section 5, we derive the equilibrium equation by using
the principle of maximizing the total energy dissipation. Then, we
construct the analytical solutions for an illustrative example. In
section 6, we further study the instability phenomena induced by
phase fronts coalescence. We try to give some descriptions and
explanations for the origin of the instability during the
coalescence process, the accompanying stress drop/jump and the
morphology varies of the specimen. We also consider the size-effect
of the specimens on the coalescence process. Finally, some
conclusions are drawn.


\section{Preliminaries}

In this section, we shall give a simple introduction to the
formulation of Rajagopal $\&$ Srinivasa (1999, 2004), which is the
starting point of our present model.

First, based on the balance laws and the local form of the entropy
production equation, one can obtain the reduced energy-rate of
dissipation relation as
$$
-(\dot{\hat{\Psi}}+\dot{T}\eta)+\textrm{tr}(\mathbf{\Sigma}
\dot{\mathbf{F}})=\zeta. \ \ \ \eqno(2.1)
$$
In equation (2.1), $\hat{\Psi}$ is the Helmholtz free energy per
unit referential volume, $T$ is the absolute temperature, $\eta$ is
the entropy per unit volume in the reference configuration and
$\zeta$ is the rate of mechanical dissipation, $\mathbf{F}$ is the
deformation gradient tensor and $\mathbf{\Sigma}$ is the nominal
stress tensor. The superposed dot indicates the material time
derivative.

To describe the phase transition process, one also needs to adopt
the phase state variable $\alpha$, which represents the volume
fraction of the martensite phase in the reference configuration. It
is assumed that $\hat{\Psi}$, $\mathbf{\Sigma}$ and $\eta$ depend on
the variables $\{\mathbf{F},\alpha,T\}$, while the rate of
mechanical dissipation $\zeta$ depends on
$\{T,\alpha,\dot{\alpha}\}$. With these constitutive assumptions,
the reduced energy-rate of dissipation relation (2.1) can be
rewritten as
$$
\textrm{tr}((\mathbf{\Sigma}-\frac{\partial \hat{\Psi}}{\partial
\mathbf{F}})\dot{\mathbf{F}})-(\frac{\partial \hat{\Psi}}{\partial
T}+\eta)\dot{T}-\frac{\partial \hat{\Psi}}{\partial
\alpha}\dot{\alpha}=\zeta, \eqno(2.2)
$$
Equation (2.2) must hold for all $\dot{\mathbf{F}}$ and $\dot{T}$,
thus one can arrive at the following constitutive equations:
$$
\mathbf{\Sigma}=\frac{\partial
\hat{\Psi}(\mathbf{F},\alpha,T)}{\partial \mathbf{F}},\ \ \eqno(2.3)
$$
$$
\eta=-\frac{\partial \hat{\Psi}(\mathbf{F},\alpha,T)}{\partial T},\
\ \eqno(2.4)
$$
and
$$
-\frac{\partial \hat{\Psi}(\mathbf{F},\alpha,T)}{\partial
\alpha}\dot{\alpha}=\zeta.\ \ \eqno(2.5)
$$
Once the specific forms of the functions $\hat{\Psi}$ and $\zeta$
are given, the nominal stress tensor $\mathbf{\Sigma}$ and entropy
$\eta$ can be calculated from (2.3) and (2.4). Equation (2.5)
provides the equation for the evolution of phase state variable
$\alpha$. As we just consider the isothermal responses of SMA
materials in this paper, the temperature $T$ will be considered as a
given constant in the sequel.

In the paper of Rajagopal $\&$ Srinivasa (1999), the following
constitutive form of the mechanical dissipation rate $\zeta$ was
proposed
$$
\zeta=\left\{
    \begin{aligned}
      &A^+(\alpha) |\dot{\alpha}|, \ \ \  \ \ \ &\textrm{if}\ \ \dot{\alpha}> 0,\\
      &0, \ \ \  \ \ \ \ \ \ \ \ \ \ \ \ \ &\textrm{if}\ \ \dot{\alpha}=0,\\
      &A^-(\alpha) |\dot{\alpha}|, \ \ \ \ \ \ &\textrm{if}\ \ \dot{\alpha}< 0,
    \end{aligned}
    \right.
\eqno(2.6)
$$
where $A^+(\alpha)\geq 0$ and $A^-(\alpha)\geq 0$ referred to as the
forward and backward dissipative resistances, respectively. An
important consequence of (2.6) is that the response of the material
is rate-independent, i.e., the stress-strain curve is the same
irrespective of the speed with which the test is conducted.

Substituting (2.6) into (2.5) and using the fact that the resulting
equation must be valid for all admissible processes, one can arrive
at the following phase transition criteria
$$
-A^-(\alpha)<-\frac{\partial \hat{\Psi}}{\partial\alpha}<A^+(\alpha)
\Rightarrow \dot{\alpha}=0,\ \ \eqno(2.7)
$$
$$
\dot{\alpha}\neq 0 \Rightarrow -\frac{\partial
\hat{\Psi}}{\partial\alpha}=\left\{
    \begin{aligned}
      &A^+(\alpha), \ \ \ &\textrm{if}\ \ \dot{\alpha}> 0,\\
      &-A^-(\alpha), \ \ \ &\textrm{if}\ \ \dot{\alpha}< 0.
    \end{aligned}
    \right.
\eqno(2.8)
$$
Condition (2.7) represents the fact that as long as $-\partial
\hat{\Psi}/\partial\alpha$ lies between $-A^-$ and $A^+$, no phase
transition can take place. On the other hand, whenever
$\dot{\alpha}\neq0$, $-\partial \hat{\Psi}/\partial\alpha$ must
equal to one of the dissipative resistances. At the two critical
points $-\partial \hat{\Psi}/\partial\alpha=A^+$ or $-\partial
\hat{\Psi}/\partial\alpha=-A^-$, the response of material may not be
unique (Rajagopal $\&$ Srinivasa 1999, 2004). To fully determine the
commencement and cessation of the phase transitions, one need to use
the criterion of maximum rate of dissipation (Rajagopal $\&$
Srinivasa 1998, 1999). Intuitively speaking, the maximum rate of
dissipation criterion states that if the material is capable of
responding in many different modes with different rates of
dissipations (including a mode that is non-dissipative), then the
actual mode will be the one with the maximum rate of dissipation.

In general, there does not exist a one-to-one mapping between
$\mathbf{F}$ and $\alpha$. Based on the phase transition criteria
and the criterion of maximum rate of dissipation, we find that for a
given deformation gradient $\mathbf{F}$, there may exist two phase
state values $\alpha^+_\mathbf{F}$ and $\alpha^-_\mathbf{F}$ such
that
$-\partial\hat{\Psi}(\mathbf{F},\alpha^+_\mathbf{F})/\partial\alpha=
A^{+}(\alpha^+_\mathbf{F})$ and
$-\partial\hat{\Psi}(\mathbf{F},\alpha^-_\mathbf{F})/\partial\alpha=-
A^{-}(\alpha^-_\mathbf{F})$, then the actual phase state $\alpha$
should satisfy $\alpha^+_\mathbf{F}\leq \alpha \leq
\alpha^-_\mathbf{F}$.

In this paper, we shall only consider two kinds of loading patterns:
the purely loading process with the initial state of the cylinder
composed of austenite phase and the purely unloading process with
the initial state of the cylinder composed of martensite phase. In
the figure of the stress-strain response, the curves corresponding
to these two kinds of loading patterns form the entire `outer loop'.

From the experiments, it was found that the deformation process at
any material point can be divided into the elastic deformation
process of the austenite phase, the elastic deformation process of
the martensite phase and the phase transition process. The
experimental results also reveal that the profile of the phase
transition (localization) region has a stable form during the phase
transition process. Thus, it is reasonable to impose the following
assumption:

\emph{Assumption 1}. In a quasi-static purely loading or purely
unloading process, for the material points located in the phase
transition region, it is satisfied that $\dot{\alpha}>0$ (for the
loading process) or $\dot{\alpha}<0$ (for the unloading process).

With this assumption, we can propose the following relationship
between $\alpha$ and $\mathbf{F}$ for the purely loading and
unloading processes, respectively:
$$
\alpha^+(\mathbf{F})=\left\{
    \begin{aligned}
      &0, \ \ \ \ \ \ \ \ \ \ \ \ \ \textrm{if}\ \ \ -\frac{\partial\hat{\Psi}(\mathbf{F},\alpha)}{\partial\alpha}|_{\alpha=0}\leq A^+(0),\\
      &1, \ \ \ \ \ \ \ \ \ \ \ \ \ \textrm{if}\ \ \ -\frac{\partial\hat{\Psi}(\mathbf{F},\alpha)}{\partial\alpha}|_{\alpha=1}\geq A^+(1),\\
      &\alpha^+_{\mathbf{F}}, \ \ \ \ \ \ \ \ \ \ \ \textrm{if}\ \ \exists\ \alpha^+_{\mathbf{F}} \in (0,1), \ \ \textrm{s.t.},\ \
      -\frac{\partial\hat{\Psi}(\mathbf{F},\alpha)}{\partial\alpha}|_{\alpha=\alpha^+_{\mathbf{F}}}=A^+(\alpha^+_{\mathbf{F}}).
    \end{aligned}
    \right.
\eqno(2.9)
$$
$$
\alpha^-(\mathbf{F})=\left\{
    \begin{aligned}
      &0, \ \ \ \ \ \ \ \ \ \ \ \ \ \textrm{if}\ \ \ -\frac{\partial\hat{\Psi}(\mathbf{F},\alpha)}{\partial\alpha}|_{\alpha=0}\leq -A^-(0),\\
      &1, \ \ \ \ \ \ \ \ \ \ \ \ \ \textrm{if}\ \ \ -\frac{\partial\hat{\Psi}(\mathbf{F},\alpha)}{\partial\alpha}|_{\alpha=1}\geq -A^-(1),\\
      &\alpha^-_{\mathbf{F}}, \ \ \ \ \ \ \ \ \ \ \ \textrm{if}\ \ \exists\ \alpha^-_{\mathbf{F}} \in (0,1), \ \ \textrm{s.t.},\ \
      -\frac{\partial\hat{\Psi}(\mathbf{F},\alpha)}{\partial\alpha}|_{\alpha=\alpha^-_{\mathbf{F}}}=-A^-(\alpha^-_{\mathbf{F}}).
    \end{aligned}
    \right.
\eqno(2.10)
$$


\section{The total elastic potential energy}

In this section, we shall derive an ``effective'' one-dimensional
expression for the total elastic potential energy of the layer,
which takes into account the higher-dimensional effects.

We consider the symmetric deformation of a thin SMA layer subject to
a static axial force at two ends. This layer can be considered as a
cross-section cutting along the axial direction of a thin SMA strip
or tube. Although here we just consider a very simple model as we
treat this layer as a two-dimensional object, it will be seen that
this simple model can capture some important experimental features.

Assume that in the stress-free configuration, the layer has
thickness $2a$ and length $l$, where $a/l=\delta\ll1$. We denote
$(X,Y)$ and $(x,y)$ the coordinates of a material point of the layer
in the reference and current configurations, respectively. The
finite displacements can be written as
$$
U(X,Y)=x(X,Y)-X,\ \ \ W(X,Y)=y(X,Y)-Y.\ \ \eqno(3.1)
$$
As here we only consider the symmetric deformation of this layer,
thus $U$ and $V$ have the following properties
$$
U(X,-Y)=U(X,Y), \ \ \ W(X,-Y)=-W(X,Y).
$$
Then the deformation gradient tensor $\mathbf{F}$ is given by
$$
\mathbf{F}=(1+U_X)\mathbf{e}_1\otimes\mathbf{E}_1+U_Y
\mathbf{e}_1\otimes \mathbf{E}_2+W_X \mathbf{e}_2 \otimes
\mathbf{E}_1+(1+W_Y) \mathbf{e}_2 \otimes \mathbf{E}_2,\ \
\eqno(3.2)
$$
where $\{\mathbf{E}_i\}_{i=1,2}$ and $\{\mathbf{e}_j\}_{j=1,2}$ are
the orthonormal base vectors in the reference and current
configurations.

Based on (2.3), (2.9) and (2.10), we obtain that the nominal stress
tensor $\mathbf{\Sigma}$ for the purely loading or purely unloading
process should be given by
$$
\mathbf{\Sigma}^{\pm}=\frac{\partial
\hat{\Psi}(\textbf{F},\alpha)}{\partial
\textbf{F}}|_{\alpha=\alpha^{\pm}(\textbf{F})}=:\textbf{G}^{\pm}(\textbf{F}),
\ \ \eqno(3.3)
$$
where $\mathbf{G}^{\pm}$ are tensor-valued functions. It will be
assumed that there exist some ``effective" strain energy functions
$\Psi^{\pm}(\mathbf{F})$ such that
$$
\mathbf{\Sigma}^{\pm}=\mathbf{G}^{\pm}(\mathbf{F})=\frac{\partial
\Psi^{\pm}(\mathbf{F})}{\partial \mathbf{F}}.\ \ \ \eqno(3.4)
$$

In this section, the purpose is to derive an ``effective''
one-dimensional expression for the elastic potential energy of the
layer. Thus, we only consider the nondissipative case in this
section, i.e., we assume $A^{\pm}(\alpha)=0$. From (2.9) and (2.10),
it is easy to see that in the nondissipative case, the phase state
functions satisfy
$\alpha^{+}(\mathbf{F})=\alpha^{-}(\mathbf{F})=:\alpha(\mathbf{F})$.
Further by using (2.9), (2.10) and (3.3), we have
$$
\begin{aligned}
\mathbf{\Sigma}(\mathbf{F})&=\frac{\partial \hat{\Psi}}{\partial
\mathbf{F}}(\mathbf{F},\alpha)|_{\alpha=\alpha(\mathbf{F})}\\&=\frac{\partial
\hat{\Psi}}{\partial
\mathbf{F}}(\mathbf{F},\alpha(\mathbf{F}))-\frac{\partial
\hat{\Psi}}{\partial
\alpha}(\mathbf{F},\alpha)|_{\alpha=\alpha(\mathbf{F})}\cdot\frac{\partial\alpha(\mathbf{F})}{\partial
\mathbf{F}}\\&=\frac{\partial
\hat{\Psi}(\mathbf{F},\alpha(\mathbf{F}))}{\partial \mathbf{F}}.
\end{aligned}
\eqno(3.5)
$$
The final equation in (3.5) is due to the fact that if $\mathbf{F}$
takes value in the phase transition domain, we have $-\frac{\partial
\hat{\Psi}}{\partial
\alpha}(\mathbf{F},\alpha)|_{\alpha=\alpha(\mathbf{F})}=0$,
otherwise $\alpha(\mathbf{F})$ should be a constant function of
$\mathbf{F}$.

From (3.5), we can see that the ``effective" strain energy function
in the nondissipative case should be given by
$$
\Psi(\mathbf{F})=\hat{\Psi}(\mathbf{F},\alpha(\mathbf{F})).
\eqno(3.6)
$$
In the sequel, we refer $\Psi(\mathbf{F})$ as the elastic potential
energy function. To go further, we propose another important
assumption:

\emph{Assumption 2}. In the nondissipative case, SMA material can be
considered as some kind of isotropic hyperelastic material. In this
case, the elastic potential energy $\Psi(\mathbf{F})$ only depends
on the two principle stretches $\lambda_1$, $\lambda_2$ of
$\mathbf{F}$; that is $\Psi=\Psi(\lambda_1,\lambda_2)$.

If the components of $\mathbf{F}-\mathbf{I}$ are relatively small,
it is possible to expand the nominal stress components in term of
the strains up to any order. The formula containing terms up to the
third order material nonlinearity is (cf. Fu $\&$ Ogden 1999)
$$
\Sigma_{ji}=a_{jilk}^1\kappa_{kl}+\frac{1}{2}a^2_{jilknm}\kappa_{kl}\kappa_{mn}+\frac{1}{6}a^3_{jilknmqp}\kappa_{kl}\kappa_{mn}\kappa_{pq}+O(|\kappa_{st}|^4),\
\ \eqno(3.7)
$$
where $\kappa_{ij}$ is the components of the tensor
$\mathbf{F}-\mathbf{I}$ and
$$
\begin{aligned}
&a_{jilk}^1=\frac{\partial^2{\Psi}}{\partial F_{ij}\partial
F_{kl}}|_{\mathbf{F}=\mathbf{I}},\ \ \ \
a_{jilknm}^2=\frac{\partial^3{\Psi}}{\partial
F_{ij}\partial F_{kl}\partial F_{mn}}|_{\mathbf{F}=\mathbf{I}},\\
&a^3_{jilknmqp}=\frac{\partial^4{\Psi}}{\partial F_{ij}\partial
F_{kl}\partial F_{mn}\partial F_{pq}}|_{\mathbf{F}=\mathbf{I}}
\end{aligned}
$$
are incremental elastic moduli, which can be calculated once a
specific form of elastic potential energy is given. From the formula
(3.7), we can obtain the nominal stress components $\Sigma_{ji}$.
For example,
$$
\begin{aligned}
\Sigma_{11}=&\tau_2W_Y+\tau_1U_X\\&+\frac{1}{2}(\eta_3U_Y^2+\eta_2W_Y^2+2\eta_2W_YU_X+\eta_1U_X^2+2\eta_4U_YW_X
+\eta_3W_X^2)\\&+\frac{1}{6}(3\theta_3U_Y^2W_Y+\theta_2W_Y^3+3\theta_3U_Y^2U_X+3\theta_5W_Y^2U_X+3\theta_2W_YU_X^2\\&
+\theta_1U_X^3+6\theta_7U_YW_YU_X+6\theta_4U_YU_XW_X+3\theta_6W_YW_X^2+3\theta_3U_XW_X^2),
\end{aligned}
\eqno(3.8)
$$
where $\tau_i$, $\eta_j$ and $\theta_k$ are some elastic moduli,
whose formulas are given in appendix A. Owing to the complexity of
calculations, we shall only work up to the third-order material
nonlinearity. In the experiments, the maximum strain is less than
$10\%$, and such an approximation is accurate enough, certainly not
worse than the trilinear approximation adopted in many works.

For a static equilibrium configuration, the nominal stress tensor
$\mathbf{\Sigma}$ satisfies the field equations
$$
\textrm{Div}(\mathbf{\Sigma})=0,
$$
which yields the following two equations
$$
\frac{\partial \Sigma_{11}}{\partial X}+\frac{\partial
\Sigma_{21}}{\partial Y}=0,\ \ \eqno(3.9)
$$
$$
\frac{\partial \Sigma_{12}}{\partial X}+\frac{\partial
\Sigma_{22}}{\partial Y}=0.\ \ \eqno(3.10)
$$
We consider the traction-free boundary conditions at $Y=\pm a$,
i.e.,
$$
\Sigma_{21}|_{Y=\pm a}=0,\ \ \ \ \Sigma_{22}|_{Y=\pm a}=0.\ \
\eqno(3.11)
$$
Equations (3.9) and (3.10) together with (3.11) provide the
governing equations for two unknowns $U$ and $W$.

We can also expand the elastic potential energy $\Psi$ in terms of
the strains up to any order. The formula containing terms up to the
fourth-order nonlinearity is given by
$$
\begin{aligned}
\Psi=&\frac{1}{2}(\tau_3U_Y^2+\tau_1W_Y^2+2\tau_2W_YU_X+\tau_1U_X^2+2\tau_3U_YW_X+\tau_3W_X^2)
\\&+\frac{1}{6}(3\eta_3U_Y^2W_Y+\eta_1W_Y^3+3\eta_3U_Y^2U_X+3\eta_2W_Y^2U_X+3\eta_2W_YU_X^2
\\&+\eta_1U_X^3+6\eta_4U_YW_YW_X+6\eta_4U_YU_XW_X+3\eta_3W_YW_X^2+3\eta_3U_XW_X^2)
\\&+\frac{1}{24}(\theta_8U_Y^4+6\theta_3U_Y^2W_Y^2+\theta_1W_Y^4+12\theta_6U_Y^2W_YU_X+4\theta_2W_Y^3U_X
\\&+6\theta_3U_Y^2U_X^2+6\theta_5W_Y^2U_X^2+4\theta_2W_YU_X^3+\theta_1U_X^4+4\theta_9U_Y^3W_X
\\&+12\theta_4U_YW_Y^2W_X+24\theta_7U_YW_YU_XW_X+12\theta_4U_YU_X^2W_X+6\theta_{10}U_Y^2W_X^2
\\&+6\theta_3W_Y^2W_X^2+12\theta_6W_YU_XW_X^2+6\theta_3U_X^2W_X^2+4\theta_9U_YW_X^3+\theta_8W_X^4).
\end{aligned}
\eqno(3.12)
$$
It can be seen that (3.12) has a very complex form. To obtain the
total elastic potential energy, one need to calculate the
integration of (3.12) over the whole layer, which will become more
complex.

Here, we shall adopt a novel approach involving coupled
series-asymptotic expansions to obtain an asymptotic expression of
the total elastic potential energy of the layer. A similar
methodology has been developed to study nonlinear waves and phase
transitions in incompressible materials (see Dai $\&$ Huo 2002, Dai
$\&$ Fan 2004, Dai $\&$ Cai 2006, Cai $\&$ Dai 2006).

Based on the symmetry of the problem (see the equation below (3.1)),
we first introduce the important transformation
$$
W(X,Y)=Y w(X,Y),\ \ \ s=Y^2.\ \eqno(3.13)
$$
From the properties of $U$ and $W$, it is easy to see that $U$ and
$w$ can be considered as functions of the variable $s$. Then we
adopt the scalings:
$$
s=l^2\tilde{s},\ \ \ X=l\tilde{x},\ \ \ U=h\tilde{u},\ \ \
w=\frac{h}{l}\tilde{w},\ \ \ \epsilon=\frac{h}{l},\ \
\nu=\frac{a^2}{l^2},\ \ \ \eqno(3.14)
$$
where $h$ is a characteristic axial displacement, and $\epsilon$
(equivalent to a small engineering strain) and $\nu$ (square of the
half thickness-length ratio) are regarded to be two small
parameters.

Substituting $(3.13)$ and $(3.14)$ into (3.9) and (3.10), we obtain
$$
\begin{aligned}
&{2 {\tau_3} u_s+({\tau_2}+{\tau_3}) w_x+{\tau_1} u_{xx}}+s (4
{\tau_3} u_{ss}+(2 {\tau_2}+2 {\tau_3}) w_{xs})+\cdots=0,
\end{aligned}
\eqno(3.15)
$$
$$
\begin{aligned}
& 6 \tau_1 w_s+(2\tau_2+2\tau_3) u_{xs}+\tau_3w_{xx}+s 4\tau_1
w_{ss}+\cdots=0.
\end{aligned}
\eqno(3.16)
$$
Here and hereafter, we have dropped the tilde for convenience. The
full forms of (3.15) and (3.16) are very lengthy and can be found in
appendix B. Here we just present the first free terms. Substituting
$(3.13)$ and $(3.14)$ into the traction-free boundary conditions
(3.11), we obtain
$$
\begin{aligned}
&2 \tau_3 u_s+\tau_3 w_x+\epsilon [2 \eta_3 w u_s+2 \eta_3 u_s
u_x+\eta_4 w w_x+\eta_4 u_x w_x\\& +s (4 \eta_3 u_s w_s+2 \eta_4 w_s
w_x)]+\epsilon ^2 [\theta_3 w^2 u_s+2 \theta_6 w u_s u_x+\theta_3
u_s u_x^2\\&+\frac{1}{2} \theta_4 w^2 w_x+\theta_7 w u_x
w_x+\frac{1}{2} \theta_4 u_x^2 w_x + s^2 (4 \theta_3 u_s w_s^2+2
\theta_4 w_s^2 w_x)\\&+ s (\frac{4}{3} \theta_8 u_s^3+4 \theta_3 w
u_s w_s+4 \theta_6 u_s w_s u_x+2 \theta_9 u_s^2 w_x+2 \theta_4 w w_s
w_x\\&+2 \theta_7 w_s u_x w_x+ \theta_{10} u_s w_x^2+\frac{1}{6}
\theta_9 w_x^3)]|_{s=\nu}=0,
\end{aligned}
\eqno(3.17)
$$
$$
\begin{aligned}
&\tau_1 w+\tau_2 u_x+ s 2 \tau_1 w_s +\epsilon[\frac{1}{2} \eta_1
w^2+s^2 2 \eta_1 w_s^2+\eta_2 w u_x+\frac{1}{2} \eta_2 u_x^2\\&+ s(2
\eta_3 u_s^2+2 \eta_1 w w_s+ 2 \eta_2 w_s u_x+2 \eta_4 u_s w_x
+\frac{1}{2} \eta_3 w_x^2)]\\&+\epsilon^2[\frac{1}{6}\theta_1
w^3+s^3 \frac{4}{3} \theta_1
w_s^3+\frac{1}{2}\theta_2w^2u_x+\frac{1}{2}\theta_5 w
u_x^2+\frac{1}{6} \theta_2 u_x^3\\&+s^2(4 \theta_3u_s^2w_s+2
\theta_1 w w_s^2 +2\theta_2 w_s^2+4
\theta_4u_sw_sw_x+\theta_3w_sw_x^2)\\&+s(2\theta_3wu_s^2+\theta_1w^2w_s+2\theta_6u_s^2u_x+2\theta_2
w w_s u_x+\theta_5 w_s u_x^2+ \\&2 \theta_4 w u_sw_x+2\theta_7
u_su_xw_x+\frac{1}{2}\theta_3ww_x^2+\frac{1}{2}\theta_6u_xw_x^2)]|_{s=\nu}=0.
\end{aligned}
\eqno(3.18)
$$

In the nonlinear PDE system (3.15)-(3.18), the two unknowns $u$ and
$w$ depend on the variable $x$, the small variable $s$ and the small
parameters $\epsilon$ and $\nu$. As long as we assume that $u$ and
$w$ are smooth enough in $s$, we can expand them in terms of the
small variable $s$:
$$
u(x,s;\epsilon,\nu)=U_0(x;\epsilon,\nu)+sU_1(x;\epsilon,\nu)+s^2U_2(x;\epsilon,\nu)+\cdots,\
\ \eqno(3.19)
$$
$$
w(x,s;\epsilon,\nu)=W_0(x;\epsilon,\nu)+sW_1(x;\epsilon,\nu)+s^2W_2(x;\epsilon,\nu)+\cdots.\
\ \eqno(3.20)
$$
Substituting (3.19) and (3.20) into the traction-free boundary
conditions (3.17) and (3.18), then by neglecting the terms higher
than $O(\epsilon\nu,\epsilon^2)$, we obtain
$$
\begin{aligned}
&2\tau_3U_1+\tau_3W_{0x}+\epsilon (2\eta_3 U_1 W_0 +2\eta_3U_1
U_{0x} +\eta_4 W_0 W_{0x}+\eta_4 U_{0x}W_{0x})\\&+\nu (4\tau_3
U_2+\tau_3 W_{1x})+\epsilon^2 (\theta_3 U_1W_0^2+2\theta_6
U_1W_0U_{0x}+\theta_3 U_1 U_{0x}^2\\& +\frac{1}{2}\theta_4
W_0^2W_{0x}+\theta_7
W_0U_{0x}W_{0x}+\frac{1}{2}\theta_4U_{0x}^2W_{0x})+\epsilon \nu
(4\eta_3 U_2W_0 \\&+6\eta_3 U_1 W_1 +4\eta_3 U_2U_{0x}+2\eta_3 U_1
U_{1x}+3\eta_4W_1W_{0x}+\eta_4U_{1x}W_{0x}\\&+\eta_4W_0W_{1x}+\eta_4
U_{0x}W_{1x})=0,
\end{aligned}
\eqno(3.21)
$$
$$
\begin{aligned}
&\tau_1W_0+\tau_2U_{0x}+\epsilon(\frac{1}{2}\eta_1W_0^2+\eta_2W_0U_{0x}+\frac{1}{2}\eta_2U_{0x}^2)+\nu
(3 \tau_1 W_1\\&+\tau_2 U_{1x}) + \epsilon^2(\frac{1}{6}\theta_1
W_0^3+\frac{1}{2}\theta_2 W_0^2U_{0x}+\frac{1}{2}\theta_5 W_0
U_{0x}^2+\frac{1}{6} \theta_2 U_{0x}^3)\\& + \epsilon \nu (2\eta_3
U_1^2 + 3 \eta_1 W_0W_1 +3\eta_2 W_1 U_{0x} +\eta_2 W_0 U_{1x}
+\eta_2 U_{0x} U_{1x} \\&+2 \eta_4 U_1 W_{0x} + \frac{1}{2} \eta_3
W_{0x}^2)=0.
\end{aligned}
\eqno(3.22)
$$
It can be seen that (3.21) and (3.22) involve the five unknowns:
$U_0$, $W_0$, $U_1$, $W_1$ and $U_2$. Thus, to form a closed-system,
we need to find another three equations which contain and only
contain these five unknowns.

Substituting (3.19) and (3.20) into (3.15), the left hand side
becomes a series in $s$ and all the coefficients of $s^n$ ($n$=$0$,
$1$, $2$, $\cdots$) should vanish. Equating the coefficients of
$s^0$ and $s^1$ to be zero yield that
$$
\begin{aligned}
&2\tau_3
U_1+(\tau_2+\tau_3)W_{0x}+\tau_1U_{0xx}+\epsilon(2\eta_3U_1W_0+2\eta_3U_1U_{0x}\\&
+(\eta_2+\eta_4)W_0W_{0x}+(\eta_2+\eta_4)U_{0z}W_{0z}+\eta_2W_0U_{0xx}+\eta_1U_{0x}U_{0xx})\\&+
\epsilon^2(\theta_3U_1W_0^2+2\theta_6U_1W_0U_{0x}+\theta_3U_1U_{0x}^2+(\frac{\theta_2}{2}+\frac{\theta_4}{2})W_0^2W_{0x}
\\&+(\theta_5+\theta_7)W_0U_{0x}W_{0x}+(\frac{\theta_2}{2}+\frac{\theta_4}{2})U_{0x}^2W_{0x}+\frac{1}{2}\theta_5W_0^2U_{0xx}\\&+\theta_2W_0U_{0x}U_{0xx}+\frac{1}{2}\theta_1U_{0x}^2U_{0xx})=0,
\end{aligned}
\eqno(3.23)
$$
$$
\begin{aligned}
&12 {\tau_3} U_2+(3 {\tau_2}+3 {\tau_3}) W_{1x}+{\tau_1}
U_{1xx}+\epsilon  (12 {\eta_3} U_2 W_0+18 {\eta_3} U_1 W_1
\\&
+12
{\eta_3} U_2 U_{0x}+10 {\eta_3} U_1
U_{1x}+(3 {\eta_2}+9 {\eta_4}) W_1 W_{0x}+({\eta 2}+5 {\eta_4})
U_{1x} W_{0x}
\\&
+(3 {\eta_2}+3 {\eta_4}) W_0 W_{1x} +(3 {\eta_2}+3 {\eta_4})
U_{0x} W_{1x}+3 {\eta_2} W_1 U_{0xx}+{\eta_1} U_{1x}
U_{0xx} \\&+{\eta_2} W_0 U_{1xx}+{\eta_1} U_{0x} U_{1xx} +2 {\eta_4}
U_1 W_{0xx}+{\eta_3} W_{0x} W_{0xx})+ {{\epsilon }^2} (4
{\theta_8} {{U_1}^3}  \\&+6 {\theta_3} U_2 {{W_0}^2}+18 {\theta_3} U_1
W_0 W_1+12 {\theta_6} U_2 W_0 U_{0x}+18 {\theta_6} U_1 W_1
U_{0x}\\&+6 {\theta_3} U_2 {{U_{0x}}^2}+10 {\theta_6} U_1 W_0 U_{1x}+10
{\theta_3} U_1 U_{0x} U_{1x}+
 (2 {\theta_6}+6 {\theta
9}) {{U_1}^2} W_{0x}\\&+(3 {\theta_2}+9 {\theta 4}) W_0 W_1
W_{0x}+(3 {\theta_5}+9 {\theta 7}) W_1 U_{0x} W_{0x}+({\theta_5}+5
{\theta_7}) W_0 U_{1x} W_{0x}\\
\end{aligned}
\eqno(3.24)
$$
$$
\begin{aligned}
&+({\theta_2}+5 {\theta_4}) U_{0x} U_{1x} W_{0x}+(3 {\theta_{10}}+2
{\theta_7}) U_1 {{W_{0x}}^2}
 +(\frac{{\theta_6}}{2}+\frac{{\theta_9}}{2}) {W_{0x}^3}\\&+(\frac{3 {\theta_2}}{2}+\frac{3 {\theta_4}}{2}) {{W_0}^2}
W_{1x}+(3 {\theta_5}+3 {\theta_7}) W_0 U_{0x} W_{1x}+(\frac{3
{\theta_2}}{2}+\frac{3 {\theta_4}}{2}) {{U_{0x}}^2} W_{1x}\\
&+2 {\theta_3} {{U_1}^2} U_{0xx}+3 {\theta_5} W_0 W_1 U_{0xx}+3
{\theta_2} W_1 U_{0x} U_{0xx}+{\theta_2} W_0 U_{1x} U_{0xx}\\&
 +{\theta_1} U_{0x} U_{1x} U_{0xx}+2 {\theta_4} U_1 W_{0x}
U_{0xx}+\frac{1}{2} {\theta_3} {{W_{0x}}^2} U_{0xx}+\frac{1}{2} {\theta_5} {{W_0}^2} U_{1xx}
 \\&
+ {\theta_2} W_0 U_{0x}
U_{1xx}+\frac{1}{2} {\theta_1} {{U_{0x}}^2} U_{1xx}+2 {\theta_7} U_1
W_0 W_{0xx}+2 {\theta_4} U_1 U_{0x} W_{0xx}\\&+{\theta_6} W_0 W_{0x}
W_{0xx}+{\theta_3} U_{0x} W_{0x} W_{0xx})=0.
\end{aligned}
$$
Similarly, substituting (3.19) and (3.20) into (3.16) and equating
the coefficient of $s^0$ to be zero yields that
$$
\begin{aligned}
& 6 {\tau_1} {W_1}+(2 {\tau_2}+2 {\tau_3}) U_{1x}+{\tau_3} W_{0xx}+
{{\epsilon }} (4 {\eta_3} {{{U_1}}^2}+6 {\eta_1} {W_0}
{W_1}\\&+(2 {\eta_2}+2 {\eta_4}) {W_0} U_{1x}+(2 {\eta_2}+2
{\eta_4}) U_{0x} U_{1x}+6 {\eta_4} {U_1} W_{0x}+2 {\eta_3}
W_{0x}^2\\&+2 {\eta_4} {U_1} U_{0xx}+{\eta_3} W_{0x} U_{0xx}+
 {\eta_3} {W_0} W_{0xx}+{\eta_3}
U_{0x} W_{0xx}+6 {\eta_2} {W_1} U_{0x})\\&+
 {{\epsilon }^2} (4
{\theta_3} {{{U_1}}^2} {W_0}+3 {\theta_1} {{{W_0}}^2} {W_1}+4
{\theta_6} {{{U_1}}^2} U_{0x}+6 {\theta_2} {W_0} {W_1} U_{0x}+3
{\theta_5} {W_1} U_{0x}^2\\&+({\theta_2}+{\theta_4}) {{{W_0}}^2}
U_{1x}+(2 {\theta_5}+2 {\theta_7}) {W_0} U_{0x}
U_{1x}+({\theta_2}+{\theta_4}) {U_{0x}^2} U_{1x} \\&+6 {\theta_4}
{U_1} {W_0} W_{0x}
 +6 {\theta_7} {U_1}
U_{0x} W_{0x}+2 {\theta_3} {W_0} {W_{0x}^2}+2 {\theta_6} U_{0x}
{W_{0x}^2} \\&+2 {\theta_7} {U_1} {W_0} U_{0xx}+2 {\theta_4} {U_1}
U_{0x} U_{0xx}+
 {\theta_6} {W_0}
W_{0x} U_{0xx}+{\theta_3} U_{0x} W_{0x} U_{0xx}\\&+\frac{1}{2}
{\theta_3} {{{W_0}}^2} W_{0xx}+{\theta_6} {W_0} U_{0x}
W_{0xx}+\frac{1}{2} {\theta_3} {U_{0x}^2} W_{0xx})=0.
\end{aligned}
\eqno(3.25)
$$

Now the nonlinear PDE system (3.15)-(3.18) has been changed into a
one-dimensional system of differential equations (3.21)-(3.25) for
the unknowns $U_0$, $W_0$, $U_1$, $W_1$ and $U_2$. Based on
(3.23)-(3.25) and by using a regular perturbation method, we can
express $U_1$, $W_1$ and $U_2$ in terms of $U_0$ and $W_0$. The
results are given below:
$$
\begin{aligned}
U_1=&(-\frac{1}{2}-\frac{\tau_2}{2\tau_3})W_{0x}-\frac{\tau_1}{2\tau_3}U_{0xx}+\epsilon
(a_1
W_0W_{0x}+a_2U_{0x}W_{0x}\\&+a_3W_0U_{0xx}+a_4U_{0x}U_{0xx})+\epsilon^2
(a_5
W_0^2W_{0x}+a_6W_0U_{0x}W_{0x}\\&+a_7U_{0x}^2W_{0x}+a_8W_0^2U_{0xx}+a_9W_0U_{0x}U_{0xx}+a_{10}U_{0x}^2U_{0xx}),
\end{aligned}
\eqno(3.26)
$$
$$
\begin{aligned}
W_1=&(\frac{\tau_2}{3\tau_1}+\frac{\tau_2^2}{6\tau_1\tau_3})W_{0xx}+(\frac{1}{6}+\frac{\tau_2}{6\tau_3})U_{0xxx}\\&+
\epsilon
(a_{11}W_{0x}^2+a_{12}W_{0x}U_{0xx}+a_{13}U_{0xx}^2+a_{14}W_0W_{0xx}\\
&+a_{15}U_{0x}W_{0xx}+a_{16}W_0U_{0xxx}+a_{17}U_{0x}U_{0xxx}),
\end{aligned}
\eqno(3.27)
$$
$$
\begin{aligned}
U_2=&(-\frac{\tau_2}{12\tau_1}+\frac{\tau_1\tau_2}{24\tau_3^2}-\frac{\tau_2^3}{24\tau_1\tau_3^2}+\frac{\tau_1}{24\tau_3}-\frac{\tau_2^2}{8\tau_1\tau_3})W_{0xxx}
\\&+(-\frac{1}{24}+\frac{\tau_1^2}{24\tau_3^2}-\frac{\tau_2^2}{24\tau_3^2}-\frac{\tau_2}{12\tau_3})U_{0xxxx}+\epsilon
(a_{18}W_{0x}W_{0xx}\\&+a_{19}U_{0xx}W_{0xx}+a_{20}W_{0x}U_{0xxx}+a_{21}U_{0xx}U_{0xxx}+a_{22}W_0W_{0xxx}\\&+a_{23}U_{0x}W_{0xxx}+a_{24}W_0U_{0xxxx}+a_{25}U_{0x}U_{0xxxx}),
\end{aligned}
\eqno(3.28)
$$
where $a_i$ $(i=1, 2, \cdots, 25)$ are material constants related to
the elastic moduli, whose expressions can be found in appendix C.
Substituting $U_1$, $W_1$ and $U_2$ into (3.21) and (3.22) and
omitting the higher order terms yield the following two equations
with only two unknowns $U_0$ and $W_0$:
$$
\begin{aligned}
&-\tau_2 W_{0x}-\tau_1 U_{0xx}+ \epsilon (-\eta_2 W_0W_{0x}-\eta_2
U_{0x} W_{0x} -\eta_2 W_0 U_{0xx}-\eta_1U_{0x}U_{0xx})\\&+ \nu
(\frac{\tau_1+2\tau_2}{6}W_{0xxx}+\frac{2\tau_1+\tau_2}{6}U_{0xxxx})+
\epsilon^2 (-\frac{1}{2}\theta_2 W_0^2W_{0x}-\theta_5
W_0U_{0x}W_{0x}\\&-\frac{1}{2}\theta_2
U_{0x}^2W_{0x}-\frac{1}{2}\theta_5 W_0^2 U_{0xx}-\theta_2 W_0 U_{0x}
U_{0xx}-\frac{1}{2}\theta_1 U_{0x}^2U_{0xx})\\& + \epsilon \nu (b_1
W_{0x}W_{0xx}-b_2U_{0x}W_{0x}-b_3 W_{0x} U_{0xxx} +b_4
U_{0xx}U_{0xxx}-b_5 W_0 W_{0xxx}\\& +b_6 U_{0x}W_{0xxx}-b_7 W_0
U_{0xxxx}-b_8 U_{0x}U_{0xxxx})=0,
\end{aligned}
\eqno(3.29)
$$
$$
\begin{aligned}
&\tau_1W_0+\tau_2U_{0x}+\epsilon (\frac{1}{2}\eta_1 W_0^2 +\eta_2
W_0 U_{0x}+\frac{1}{2}\eta_2 U_{0x}^2) \\&+ \nu
(\frac{1}{2}\tau_2W_{0xx}+\frac{1}{2}\tau_1
U_{0xxx})+\epsilon^2(\frac{1}{6}\theta_1W_0^3 +\frac{1}{2} \theta_2
W_0^2 U_{0x}+\frac{1}{2}\theta_5 W_0 U_{0x}^2
\\&+\frac{1}{6}\theta_2 U_{0x}^3)+ \epsilon \nu (-b_9
W_{0x}^2-b_{10} W_{0x}U_{0xx}-b_{11}U_{0xx}^2-b_{12}W_0
W_{0xx}\\&-b_{13}U_{0x} W_{0xx}-b_{14}W_0
U_{0xxx}-b_{15}U_{0x}U_{0xxx})=0.
\end{aligned}
\eqno(3.30)
$$
Integrating (3.29) once, we obtain
$$
\begin{aligned}
&C-\tau_2W_0-\tau_1U_{0x}+\epsilon (-\frac{1}{2}\eta_2 W_0^2 -\eta_2
W_0 U_{0x}-\frac{1}{2}\eta_1 U_{0x}^2)+ \nu
(\frac{\tau_1+2\tau_2}{6}W_{0xx} \\&+\frac{2\tau_1+\tau_2}{6}
U_{0xxx})+\epsilon^2(-\frac{1}{6}\theta_2W_0^3 -\frac{1}{2} \theta_5
W_0^2 U_{0x}-\frac{1}{2}\theta_2 W_0 U_{0x}^2-\frac{1}{6}\theta_1
U_{0x}^3)
\\&+ \epsilon \nu (-b_{16}
W_{0x}^2-b_{17} W_{0x}U_{0xx}-b_{18}U_{0xx}^2-b_{19}W_0
W_{0xx}+b_{20}U_{0x} W_{0xx}\\&-b_{21}W_0
U_{0xxx}-b_{22}U_{0x}U_{0xxx})=0,
\end{aligned}
\eqno(3.31)
$$
where $C$ is the integration constant. The coefficients $b_i$ $(i=1,
2, \cdots, 22)$ in (3.29)-(3.31) are also some constants related to
the elastic moduli, whose expressions can be found in appendix D.

It is important to find the physical meaning of $C$, since to
capture the instability phenomena observed in the experiments, one
needs to study the global bifurcation as the physical parameters
vary. For that purpose, we consider the resultant force $P$ acting
on the material cross-section that is planar and perpendicular to
the X-axis in the reference configuration, and the formula is
$$
P=\int_{-a}^a \Sigma_{11} dY.\ \ \ \eqno(3.32)
$$
The formula for the nominal stress component $\Sigma_{11}$ has
already been given in (3.8). Through the transformations
(3.13)-(3.14) and the series expansions (3.19)-(3.20), we can
express $\Sigma_{11}$ in terms of $U_0$, $W_0$, $U_1$, $W_1$ and
$U_2$. Further by using (3.26)-(3.28), we get an expression of
$\Sigma_{11}$ which only depends on $U_0$ and $W_0$. Then, carrying
out the integration in (3.32), we obtain that
$$
\begin{aligned}
P=&2 a \epsilon [\tau_2W_0+\tau_1U_{0x}+\epsilon (\frac{1}{2}\eta_2
W_0^2 +\eta_2 W_0 U_{0x}+\frac{1}{2}\eta_1 U_{0x}^2)+
\\&\nu
(-\frac{\tau_1+2\tau_2}{6}W_{0xx}-\frac{2\tau_1+\tau_2}{6}
U_{0xxx})+\epsilon^2(\frac{1}{6}\theta_2W_0^3+ \frac{1}{2} \theta_5
W_0^2 U_{0x}
\\&+\frac{1}{2}\theta_2 W_0 U_{0x}^2+\frac{1}{6}\theta_1
U_{0x}^3)+ \epsilon \nu (b_{16} W_{0x}^2+b_{17}
W_{0x}U_{0xx}+b_{18}U_{0xx}^2\\&+b_{19}W_0 W_{0xx}-b_{20}U_{0x}
W_{0xx}+b_{21}W_0 U_{0xxx}+b_{22}U_{0x}U_{0xxx})].
\end{aligned}
\eqno(3.33)
$$
Comparing (3.31) and (3.33), we have $C=P/(2 a \epsilon)$.

By using the two equations (3.29) and (3.30), we can express $W_0$
in terms of $U_0$. First, from (3.29), we have
$$
\begin{aligned}
W_{0x}=\frac{1}{\tau_2}(-\tau_1U_{0xx}+\epsilon(-\eta_2W_0W_{0x}-\eta_2U_{0x}W_{0x}-\eta_2W_0U_{0xx}-\eta_1U_{0x}U_{0xx})).
\end{aligned}
\eqno(3.34)
$$
Substituting (3.34) into the term $\nu
(\frac{1}{2}\tau_2W_{0xx}+\frac{1}{2}\tau_1 U_{0xxx})$ of (3.30), we
obtain the following equation
$$
\begin{aligned}
&\tau_1W_0+\tau_2U_{0x}+\epsilon(\frac{1}{2}\eta_1W_0^2+\eta_2W_0U_{0x}+\frac{1}{2}\eta_2U_{0x}^2)
+\epsilon^2(\frac{1}{6}\theta_1W_0^3\\&+\frac{1}{2}\theta_2W_0^2U_{0x}+\frac{1}{2}\theta_5W_0U_{0x}^2+\frac{1}{6}\theta_2U_{0x}^3)
+\epsilon
\nu\frac{\tau_1(\tau_1+\tau_2)}{2(\tau_1-\tau_2)}(-W_{0x}^2-U_{0xx}^2\\&-2W_{0x}U_{0xx}-W_0W_{0xx}-U_{0x}W_{0xx}-W_0U_{0xxx}-U_{0x}U_{0xxx})=0.
\end{aligned}
\eqno(3.35)
$$
From (3.35) and by using a regular perturbation method, we obtain
$$
\begin{aligned}
W_0=&-\frac{\tau_2}{\tau_1}U_{0x}+\epsilon
\alpha_1U_{0x}^2+\epsilon^2 \alpha_2 U_{0x}^3\\& +\epsilon\nu
((\frac{1}{2}-\frac{\tau_2^2}{2\tau_1^2})U_{0xx}^2+(\frac{1}{2}-\frac{\tau_2^2}{2\tau_1^2})U_{0x}U_{0xxx}),
\end{aligned}
\eqno(3.36)
$$
where
$$
\begin{aligned}
\alpha_1=&-\frac{\tau_2^2\eta_1+\tau_1^2\eta_2-2\tau_1\tau_2\eta_2}{2\tau_1^3},\\
\alpha_2=&\frac{1}{6\tau_1^5}(\tau_2^3(-3\eta_1^2+\tau_1\theta_1)+\tau_1^3(3\eta_2^2-\tau_1\theta_2)-3\tau_1\tau_2^2(-3\eta_1\eta_2+\tau_1\theta_2)\\&+3\tau_1^2\tau_2(-\eta_1\eta_2-2\eta_2^2+\tau_1\theta_5)).
\end{aligned}
$$

By using the coupled series-asymptotic expansion method introduced
above, we can derive a one-dimensional asymptotic expression for
the total elastic potential energy of the layer. The elastic
potential energy per unit referential length of the layer is
$$
\bar{\Psi}=\int_{-a}^{a}\Psi dY. \ \ \eqno(3.37)
$$
The formula for the elastic potential energy $\Psi$ has already been
given in (3.12). Through the manipulations we just described above,
we obtain
$$
\begin{aligned}
\bar{\Psi}=&2 a \epsilon^2
[\frac{1}{2}\tau_1W_0^2+\tau_2W_0U_{0x}+\frac{1}{2}\tau_1U_{0x}^2+\epsilon(\frac{1}{6}\eta_1W_0^3+\frac{1}{2}\eta_2W_0^2U_{0x}^2\\&
+\frac{1}{2}\eta_2W_0U_{0x}^2+\frac{1}{6}\eta_1U_{0x}^2)+\nu(\frac{\tau_2^2}{3(\tau_1-\tau_2)}W_{0x}^2+\frac{2\tau_1\tau_2}{3(\tau_1-\tau_2)}W_{0x}U_{0xx}\\&
+\frac{\tau_1^2}{3(\tau_1-\tau_2)}U_{0xx}^2+\frac{1}{6}\tau_2W_0W_{0xx}+\frac{1}{6}(-\tau_1-2\tau_2)U_{0x}W_{0xx}\\&+\frac{1}{6}\tau_1W_0U_{0xxx}
+\frac{1}{6}(-2\tau_1-\tau_2)U_{0x}U_{0xxx})+\epsilon^2(\frac{1}{24}\theta_1W_0^4\\&
+\frac{1}{6}\theta_2W_0^3U_{0x}+\frac{1}{4}\theta_5W_0^2U_{0x}^2+\frac{1}{6}\theta_2W_0U_{0x}^3+\frac{1}{24}\theta_1U_{0x}^4)\\&
+\epsilon\nu(c_1W_0W_{0x}^2+c_2U_{x}W_{0x}^2+c_3W_0W_{0x}U_{0xx}+c_4U_{0x}W_{0x}U_{0xx}\\&
+c_5W_0U_{0xx}^2+c_6U_{0x}U_{0xx}^2+c_7W_0^2W_{0xx}+c_8W_0U_{0x}W_{0xx}\\&+c_9U_{0x}^2W_{0xx}+c_{10}W_0^2U_{0xxx}+c_{11}W_0U_{0x}U_{0xxx}+c_{12}U_{0x}^2U_{0xxx})]
\end{aligned}
\eqno(3.38)
$$
where $c_i$ $(i=1,\cdots,12)$ are some constants, whose expressions
can be found in appendix E. By further using (3.34) and (3.36), we
can reduce the above expression as
$$
\begin{aligned}
\bar{\Psi}=&2 a \epsilon^2 E
[\frac{1}{2}U_{0x}^2+\frac{1}{3}D_1\epsilon
U_{0x}^3+\frac{1}{4}D_2\epsilon^2U_{0x}^4+\nu(F_1 U_{0xx}^2+F_2
U_{0x}U_{0xxx})\\&+\epsilon\nu(F_3U_{0x}U_{0xx}^2+F_4U_{0x}^2U_{0xxx})],
\end{aligned}
\eqno(3.39)
$$
where $D_1$, $D_2$, $F_1$--$F_4$ are some material constants and $E$
is the Young's modulus. The formulas for these constants are given
by
$$
\begin{aligned}
D_1=&\frac{\tau_1^2\eta_1+\tau_2^2\eta_1+\tau_1\tau_2(\eta_1-3\eta_2)}{2\tau_1^2(\tau_1+\tau_2)},\\
D_2=&\frac{1}{6\tau_1^4(\tau_1^2-\tau_2^2)}[\tau_2^4(-3\eta_1^2+\tau_1\theta_1)+\tau_1^4(-3\eta_2^2+\tau_1\theta_1)-4\tau_1\tau_2^3(-3\eta_1\eta_2\\
&+\tau_1\theta_2)-4\tau_1^3\tau_2(-3\eta_2^2+\tau_1\theta_2)+6\tau_1^2\tau_2^2(-\eta_1\eta_2-2\eta_2^2+\tau_1\theta_5)],\\
E=&\frac{\tau_1^2-\tau_2^2}{\tau_1},\ \ \
F_1=-\frac{\tau_1+\tau_2}{6\tau_1},\ \ \
F_2=-\frac{2\tau_1+\tau_2}{6\tau_1},\\
F_3=&\frac{-\tau_1(\tau_2+\eta_1-\eta_2)+\tau_2(-\eta_1+\eta_2)}{6\tau_1^2},\\
F_4=&\frac{1}{12\tau_1^3(\tau_1+\tau_2)}[2\tau_2^4+4\tau_1^3(\tau_2-\eta_1)-3\tau_2^3\eta_1+\tau_1\tau_2^2(-7\eta_1+9\eta_2)\\&+\tau_1^2\tau_2(2\tau_2-7\eta_1+12\eta_2)].
\end{aligned}
$$
By retaining the original dimensional variable, we obtain
$$
\begin{aligned}
\bar{\Psi}=&2 a E\
[\frac{1}{2}V^2+\frac{1}{3}D_1V^3+\frac{1}{4}D_2V^4+a^2 (F_1
V_{X}^2+F_2VV_{XX})\\&+a^2(F_3VV_X^2+F_4V^2V_{XX})],
\end{aligned}
\eqno(3.40)
$$
where $V(X)=\epsilon \widetilde{u}_{0x}=U_{0X}$ is the leading order
term of the axial strain. Then the total elastic potential energy of
the layer should be given by
$$
\begin{aligned}
\Phi_E=\int_0^l\bar{\Psi} dX=&2 a E
\int_0^l[\frac{1}{2}V^2+\frac{1}{3}D_1V^3+\frac{1}{4}D_2V^4+a^2 (F_1
V_{X}^2\\&+F_2 VV_{XX})+a^2(F_3VV_X^2+F_4V^2V_{XX})] dX.
\end{aligned}
\eqno(3.41)
$$
Although the expression of $\Phi_E$ given in (3.41) is
one-dimensional, it is derived from a higher-dimensional setting and
has the higher-dimensional effects built in. It should be noted that
the expression (3.41) also contains higher-order derivative terms
which play the role of regularization. Trunskinovsky (1982, 1985)
(see also, Aifantis $\&$ Serrin 1983 and Triantafyllidis $\&$
Aifantis 1986) pioneered the idea of the regularization augmentation
for solid-solid phase transitions which involves adding terms (like
a strain gradient) into the usual constitutive stress-strain
relation. Here the gradient terms in our expression is also derived
and its coefficient is explicitly given.



\section{The mechanical dissipation functions}

One of the most important experimental results is that the measured
engineering stress-strain response is rate-independent hysteretic.
To model this rate-independent hysteresis phenomenon, one needs to
take into account the mechanical dissipation due to phase
transitions. In this section, we shall derive the expressions of the
mechanical dissipation functions in terms of the axial strain for
the purely loading and purely unloading processes, which can be used
to determine the total amount of energy dissipated during the phase
transitions.

The constitutive form of the mechanical dissipation rate $\zeta$ has
already been given in (2.6). From (2.6), it can be seen that the
energy dissipation only occurs whenever $\dot{\alpha}\neq0$ and also
depends on the phase state variable $\alpha$. Thus, to determine the
energy dissipation rate at certain material point, we first need to
determine the value of the phase state variable.

Based on assumption 1 and the phase transition criteria (2.7)-(2.8),
we have already proposed the evolution laws of the phase state
$\alpha$ for the purely loading and purely unloading processes in
(2.9) and (2.10).

For the purpose of simplicity, the mechanical dissipation effect
will be considered in a one-dimensional setting in this paper, i.e.,
we neglect the influence of the radial strain on the mechanical
dissipation. In this case, the deformation gradient tensor
$\mathbf{F}$ should be replaced by the axial strain $V$. Thus, the
evolution laws (2.9) and (2.10) reduce to
$$
\alpha^+(V)=\left\{
    \begin{aligned}
      &0, \ \ \ \ \ \ \ \textrm{if}\ \ \ 0\leq V <V_0^+,\\
      &1, \ \ \ \ \ \ \ \textrm{if}\ \ \ V_1^+<V \leq V^*,\\
      &\alpha^+_V,\ \ \ \ \ \textrm{if}\ \ \ V_0^+\leq V \leq V_1^+,
    \end{aligned}
    \right.
\eqno(4.1)
$$
and
$$
\alpha^-(V)=\left\{
    \begin{aligned}
      &0, \ \ \ \ \ \ \ \textrm{if}\ \ \ 0\leq V <V_0^-,\\
      &1, \ \ \ \ \ \ \ \textrm{if}\ \ \ V_1^-<V \leq V^*,\\
      &\alpha^-_V,\ \ \ \ \ \textrm{if}\ \ \ V_0^-\leq V \leq V_1^-,
    \end{aligned}
    \right.
\eqno(4.2)
$$
where the critical strains $V_0^{\pm}$ and $V_1^{\pm}$ are
determined by
$$
\begin{aligned}
&-\frac{\partial\hat{\Psi}(V_0^{+},\alpha)}{\partial\alpha}|_{\alpha=0}=
A^{+}(0), \ \ \
-\frac{\partial\hat{\Psi}(V_1^{+},\alpha)}{\partial\alpha}|_{\alpha=1}=
A^{+}(1),\\
&-\frac{\partial\hat{\Psi}(V_0^{-},\alpha)}{\partial\alpha}|_{\alpha=0}=-
A^{-}(0), \ \ \
-\frac{\partial\hat{\Psi}(V_1^{-},\alpha)}{\partial\alpha}|_{\alpha=1}=-
A^{-}(1)
\end{aligned}
\eqno(4.3)
$$
and $\alpha^{\pm}_V$ satisfy
$$
-\frac{\partial\hat{\Psi}(V,\alpha)}{\partial\alpha}|_{\alpha=\alpha^+_V}=
A^{+}(\alpha^+_V),\ \ \
-\frac{\partial\hat{\Psi}(V,\alpha)}{\partial\alpha}|_{\alpha=\alpha^-_V}=
-A^{-}(\alpha^-_V).\ \ \eqno(4.4)
$$
Here, $V^*$ denotes the maximum strain value in the experiments. It
is easy to see that $\alpha^\pm(V)$ are continuous functions. In
principle, once $\hat{\Psi}$, $A^+(\alpha)$ and $A^-(\alpha)$ are
given, we can obtain the explicit expressions of the phase state
functions $\alpha^{\pm}_V$.

In the paper of Rajagopal $\&$ Srinivasa (1999), some specific
constitutive structures were proposed for the Helmholtz free energy
$\hat{\Psi}(V,\alpha)$ and the dissipative response functions
$A^+(\alpha)$ and $A^-(\alpha)$. Through some calculations, it was
found that $\alpha^\pm(V)$ were monotonically increasing functions
during the intervals $(V_0^{\pm},V_1^{\pm})$. Based on the model
proposed by Rajagopal $\&$ Srinivasa (1999), we assume that
$\alpha^\pm(V)$ have the following property:

\emph{Assumption 3}. $\alpha^\pm(V)$ are monotonically increasing
functions during the intervals $(V_0^{\pm},V_1^{\pm})$.

By virtue of the phase state functions $\alpha^+(V)$ and
$\alpha^-(V)$, we can express the total amount of mechanical
dissipations during the purely loading and purely unloading
processes as functions of the axial strain $V(Z)$. Suppose that the
loading process starts (say, at time $t_0$) from the homogeneous
configuration $V(Z)=0$ ($0\leq Z\leq l$). Then the mechanical
dissipation (say, at time $t$) through the loading process should be
given by
$$
\begin{aligned}
\Phi^+_D&=\int_0^l\int_{-a}^a\int_{t_0}^t\zeta dtdXdY\\
&=\int_0^l\int_{-a}^{a}\int_{t_0}^tA^+(\alpha)\dot{\alpha}dtdXdY\\
&=\int_0^l
\int_{-a}^a\int_{0}^{\alpha^{+}(V(X))}A^{+}(\alpha)d\alpha
 dY dX\\
&=2 a \int_0^l \int_0^{V(X)} A^+(\alpha^+(v)) \frac{d\alpha^+(v)}{dv}dv dX\\
&=2 a E \int_0^l \phi_d^+(V(X)) dX,
\end{aligned}
\eqno(4.5)
$$
where
$$
\phi_d^+(V)=\frac{1}{E}\int_0^{V} A^+(\alpha^+(v))
\frac{d\alpha^+(v)}{dv}dv.\ \ \ \eqno(4.6)
$$

Similarly, for the unloading process (suppose that the unloading
process starts from the homogeneous configuration $V(X)=V^*>V_1^-$
($0\leq X\leq l$)), we can deduce
$$
\begin{aligned}
\Phi^-_D=2 a E \int_0^l \phi_d^-(V(X)) dX,
\end{aligned}
\eqno(4.7)
$$
where
$$
\phi_d^-(V)=-\frac{1}{E}\int_{V^*}^{V} A^-(\alpha^-(v))
\frac{d\alpha^-(v)}{dv}dv.\ \ \ \eqno(4.8)
$$
The two functions $\phi_d^+(V)$ and $\phi_d^-(V)$ defined in (4.6)
and (4.8) are referred as the dissipation density functions.

In principle, once $\hat{\Psi}$, $A^+(\alpha)$ and $A^-(\alpha)$ are
given, $\phi_d^+(V)$ and $\phi_d^-(V)$ can be obtained. We also
point out that since the strain is a measurable quantity, it may be
easier to determine $\phi_d^+(V)$ and $\phi_d^-(V)$ experimentally
than $A^+(\alpha)$ and $A^-(\alpha)$.


\section{The equilibrium equation and analytical solutions}

With the expressions of the total elastic potential energy and the
mechanical dissipation functions, we can determine the equilibrium
configurations of the layer during the phase transition process. We
shall further consider an illustrative example with some given
material constants and some special forms of dissipation functions
in this section. Subject to the free end boundary conditions at the
two ends of the layer, we construct the analytical solutions, which
can capture some important experimental features.

\subsection{Equilibrium configurations}

In this subsection, we shall determine the equilibrium
configurations of the layer during the phase transition process by
using the variational method.

In a general case, we can prove that for the purely loading and
purely unloading processes, the equilibrium configuration of the SMA
body can by determined by using the principle of maximizing total
energy dissipation (see Wang $\&$ Dai 2009). The ``total energy
dissipation" is here referred as the part of work done by the
external force that is not converted into the elastic energy or used
to overcome the mechanical dissipation due to phase transformation.

In the current case, the total energy dissipations during the purely
loading and purely unloading processes should be given by
$$
\begin{aligned}
\mathscr{E}^+(V)=&2 a E\int_0^l\gamma (V(X)-0)
dX-(\Phi_E(V)-\Phi_E(0))-\Phi^+_D(V)\\
=&2 a E\int_0^l(\gamma
V-(\frac{1}{2}V^2+\frac{1}{3}D_1V^3+\frac{1}{4}D_2V^4\\&+a^2(F_1V_X^2+F_2VV_{XX}+F_3VV_X^2+F_4V^2V_{XX}))-\phi_d^+(V))dX.
\end{aligned}
\eqno(5.1)
$$
and
$$
\begin{aligned}
\mathscr{E}^-(V)=&2 a E \int_0^l\gamma
(V(X)-V^*)dX-(\Phi_E(V)-\Phi_E^*)-\Phi^-_D(V)\\
=&\Phi_E^*-2 a E \int_0^l\gamma V^*dX\\&+2 a E\int_0^l(\gamma
V-(\frac{1}{2}V^2+\frac{1}{3}D_1V^3+\frac{1}{4}D_2V^4\\&+a^2(F_1V_X^2+F_2VV_{XX}+F_3VV_X^2+F_4V^2V_{XX}))-\phi_d^-(V))dX,
\end{aligned}
\eqno(5.2)
$$
where $\gamma=P/(2 a E)$ is the dimensionless engineering stress
acting on per unit reference area of the cross-section of the layer
and $\Phi_E^*$ represents the total elastic potential energy of the
layer corresponding to the initial state of the unloading process.

To determine the equilibrium configurations of the layer during the
phase transition process, we need to find $V(X)$ such that
$\mathscr{E}^+(V)$ and $\mathscr{E}^-(V)$ attain the maximum values.
From (5.1) and (5.2), by using the variational principle, we obtain
that the equilibrium configurations should satisfy the following
equation
$$
V+D_1V^2+D_2V^3+a^2(-\frac{1}{3}V_{XX}+D_3V_X^2+2D_3VV_{XX})+\frac{\partial
\phi_d^{\pm}(V)}{\partial V}=\gamma,\ \ \eqno(5.3)
$$
where $D_3=2F_4-F_3$. We refer (5.3) as the equilibrium equation.

\noindent\textbf{Remark}: From another point of view, the
equilibrium configurations of the layer during the phase transition
process can also be determined by using the principle of minimizing
the total ``pseudo-potential energy" of the whole mechanical system.
The concept of ``pseudo-elastic energy" function was first
introduced by Oritz $\&$ Repetto (1999) to study the dislocation
structures in plastically deformed crystals.

In our case, the pseudo-elastic energy functions
${\cal{W}}^{\pm}(V)$ corresponding to loading and unloading
processes should be defined by
$$
{\cal{W}}^{+}(V)=\Phi_E(V)+\Phi_D^+(V),\ \ \eqno(5.4)
$$
and
$$
{\cal{W}}^{-}(V)=\Phi_E(V)-\Phi_E^*+\Phi_D^-(V).\ \ \eqno(5.5)
$$

Based on the pseudo-elastic energy functions, the total
pseudo-potential energies of the whole mechanical system can be
defined by
$$
\Omega^{\pm}={\cal{W}}^{\pm} - P\int_0^lVdX. \eqno(5.6)
$$

With the expressions of the total pseudo-potential energy (5.6) and
by using the variational method, we can also derive the equilibrium
equation (5.3).

In fact, from (5.1), (5.2) and (5.6), it is easy to see that the
principle of maximizing the ``total energy dissipation'' is
equivalent to the principle of minimizing the ``total
pseudo-potential'' energy. For the purpose of convenience, we shall
only use the principle of minimizing the total pseudo-potential
energy in the sequel. Notice that both the principles of maximizing
the total energy dissipation and minimizing the total
pseudo-potential energy do not reveal anything about the path taken
by the layer from its initial configuration to the final equilibrium
configuration, i.e., these principles are entirely silent about the
process and are only concerned about the initial and final states.

\noindent\textbf{Remark}: Mielke \emph{et al.} (2002) proposed a
rate-independent, mesoscopic model for the hysteretic evolution of
phase transformations in SMAs. In their model, an extremum
principle, called the postulate of realizability (cf. Levitas 1995a,
b), was adopted to determine the stable state of the material during
the phase transformation process. We can also demonstrate that the
extremum principles we used in this section are consistent with the
postulate of realizability (cf. Wang $\&$ Dai 2009).

\subsection{The analytical solutions}

In this subsection, we shall construct the analytical solutions for
an illustrative example with some given material constants $D_1$,
$D_2$, $D_3$ and some special form of dissipation density functions
$\phi_d^{\pm}(V)$. It will be seen that the solutions obtained can
be used to explain some important experimental results. We also
point out that most of the conclusions drawn from this illustrative
example also hold in the general case.

First, without loss of generality, we take the length of the layer
to be 1. At the two ends of the layer, we impose the free end
boundary conditions. By `free ends', we mean that
$$
V_X=0, \ \ \ \textrm{at}\ X=0, 1,\ \ \ \eqno(5.7)
$$
which is sometimes called natural boundary conditions and has been
used by many authors (e.g., Ericksen 1975; Tong \emph{et al.} 2001).

Next, we shall give some further discussion on the dissipation
density functions $\phi_d^{\pm}(V)$. Based on (4.6) and (4.8) and by
considering the properties of the phase state functions
$\alpha^{\pm}(V)$, we get that $\phi_d^{\pm}(V)$ have the following
general properties:
\begin{itemize}

\item $\phi_d^+(V)$ is a continuous monotonically increasing function; $\phi_d^-(V)$ is a continuous
monotonically decreasing function.

\item $\phi_d^+(V)$ equals $0$ for $0\leq V \leq V_0^+$ and is a constant for $V_1^+\leq V
         \leq V^*$; $\phi_d^-(V)$ equals $0$ for $V_1^-\leq V
         \leq V^*$ and is a constant for $0\leq V \leq V_0^-$.

\end{itemize}

Due to the diversity of materials, the dissipation density functions
$\phi_d^{\pm}(V)$ could have many different forms. In this section,
we choose $\phi_d^{\pm}(V)$ to be fourth-order polynomials in the
intervals $(V^{\pm}_0,V_1^{\pm})$, which are given by
$$
\begin{aligned}
\phi_d^+(V)&=\bar{H}_1^+(V-V^+_0)+\bar{H}_2^+(V-V^+_0)^2+\bar{H}_3^+(V-V^+_0)^3+\bar{H}_4^+(V-V^+_0)^4\nonumber\\
&=\tilde{H}^+_0+\tilde{H}^+_1V+\tilde{H}^+_2V^2+\tilde{H}^+_3V^3+\tilde{H}^+_4V^4
\end{aligned}
\eqno(5.8)
$$
and
$$
\begin{aligned}
\phi_d^-(V)&=\bar{H}_1^-(V-V^-_1)+\bar{H}_2^-(V-V^-_1)^2+\bar{H}_3^-(V-V^-_1)^3+\bar{H}_4^-(V-V^-_1)^4\nonumber\\
&=\tilde{H}^-_0+\tilde{H}^-_1V+\tilde{H}^-_2V^2+\tilde{H}^-_3V^3+\tilde{H}^-_4V^4.
\end{aligned}
\eqno(5.9)
$$

\noindent\textbf{Remark}: Here $\phi_d^{\pm}(V)$ are chosen to be
fourth-order polynomials for the purpose of simplicity. In the
general case, if $\phi_d^{\pm}(V)$ are smooth enough, we can also
consider the Taylor series expansions of $\phi_d^{\pm}(V)$ in the
intervals $(V_0^{\pm},V_1^{\pm})$.

By substituting (5.8) and (5.9) into the equilibrium equation (5.3),
we obtain
$$
\left\{
    \begin{aligned}
      -\gamma+V+D_1V^2+D_2V^3-\frac{1}{3}a^2V_{XX}+&a^2(D_3V_X^2+2D_3VV_{XX})=0, \\ &\textrm{if}\ 0\leq V\leq V_0^{\pm} \ \ \textrm{or}\ \ V_1^{\pm}\leq V< V^*,\\
      -(\gamma-\hat{H}^{\pm}_0)+\hat{H}^{\pm}_1V+\hat{H}^{\pm}_2V^2+\hat{H}^{\pm}_3V^3-&\frac{1}{3}a^2V_{XX}+a^2(D_3V_X^2+2D_3VV_{XX})=0, \\ &\textrm{if}\ V_0^{\pm}\leq V\leq V_1^{\pm},\\
    \end{aligned}
    \right.
\eqno(5.10)
$$
where
$$
\hat{H}^{\pm}_0=\tilde{H}^{\pm}_1,\ \ \
\hat{H}^{\pm}_1=1+2\tilde{H}^{\pm}_2,\ \ \
\hat{H}^{\pm}_2=D_1+3\tilde{H}^{\pm}_3,\ \ \
\hat{H}^{\pm}_3=D_2+4\tilde{H}^{\pm}_4. \eqno(5.11)
$$

For the purpose of illustration, we choose the following numerical
values in this paper
$$
\begin{aligned}
&D_1=-23.81,\ \ D_2=158.73,\ \ D_3=-20/3,\ \ V^*=0.1 \\
&\hat{H}^+_0=0.0087,\ \ \hat{H}^+_1=0.3132,\ \ \hat{H}^+_2=-7.1069,\
\ \hat{H}^+_3=41.9287,\\
&\hat{H}^-_0=0.0097,\ \ \hat{H}^-_1=0.1497,\ \ \hat{H}^-_2=-5.4717,\
\ \hat{H}^-_3=41.9287.
\end{aligned}
\eqno(5.12)
$$
With the above chosen material constants and through some
calculations, we can get the values of some critical strains and
stresses, which are given by
$$
\begin{aligned}
&V_0^+=0.03,\ \ \ V_1^+=0.083,\ \ \ V_0^-=0.017,\ \ \ V_1^-=0.07,\\
&\xi_1^+=0.00973603,\ \ \ \xi_2^+=0.01285714,\ \ \
\xi_m^+=0.01143011,\\
&\xi_1^-=0.00777778,\ \ \ \xi_2^-=0.01089889,\ \ \
\xi_m^-=0.00920481,
\end{aligned}
\eqno(5.13)
$$
where $\xi_1^{\pm}$ are the valley stress values corresponding to
$V_1^{\pm}$, $\xi_2^{\pm}$ are the peak stress values corresponding
to $V_0^{\pm}$ and $\xi_m^{\pm}$ are the Maxwell stress values.

Next, we shall solve equation (5.10) under the free end boundary
conditions (5.7). Here, we omit the detailed derivation process,
which can be found in our previous work (see Wang $\&$ Dai 2009),
and just give the expressions for the analytical solutions.

First, we find that there exist constant solutions, which are given
by the real roots of the following equations
$$
\gamma=\left\{
         \begin{aligned}
           &V+D_1V^2+D_2V^3,\ \ \ \ \ \ 0\leq V< V_0^{\pm}\ \ \textrm{or}\ \ V_1^{\pm}<V\leq V^*,\\
           &\hat{H}^{\pm}_0+\hat{H}^{\pm}_1V+\hat{H}^{\pm}_2V^2+\hat{H}^{\pm}_3V^3,\ \ \ \ \ \ V_0^{\pm}\leq V\leq V_1^{\pm}.\ \ \ \
\end{aligned}
\right. \eqno(5.14)
$$
The constant solutions correspond to the homogeneous deformations of
the layer. It's clear that these constant solutions satisfy
$V_X\equiv0$ for $X\in[0,1]$, thus the boundary conditions (5.7) are
satisfied. It should be noted that as $\gamma$ varies, the number of
the constant solutions can also be different. In fact, there is only
one constant solution if $0\leq\gamma<\xi_1^{\pm}$ or
$\gamma>\xi_2^{\pm}$; there are two constant solutions if
$\gamma=\xi_1^{\pm}$ or $\gamma=\xi_2^{\pm}$; there are three
constant solutions if $\xi_1^{\pm}<\gamma<\xi_2^{\pm}$.

Besides the constant solutions, there are also non-trivial solutions
when $\xi_1^{\pm}<\gamma<\xi_2^{\pm}$. The expressions for the
nontrivial solutions are given by:
$$
\begin{aligned}
X&=a\int_{g_1^{\pm}}^V\sqrt{\frac{\frac{1}{6}-D_3\tau}{C^{\pm}+f^{\pm}(\tau)}}d\tau,\
\ \
V|_{X=0}=g_1^{\pm}, \\
X&=-a\int_{g_2^{\pm}}^V\sqrt{\frac{\frac{1}{6}-D_3\tau}{C^{\pm}+f^{\pm}(\tau)}}d\tau,\
\ \ V|_{X=0}=g_2^{\pm},
\end{aligned}
\eqno(5.15)
$$
where $C^{\pm}$ is an integration constant and
$$
f^{\pm}(V)=\left\{
\begin{aligned}
f_1^{\pm}(V)&=-\gamma
V+\frac{1}{2}V^2+\frac{1}{3}D_1V^3+\frac{1}{4}D_2V^4,\ \ 0\leq V<V_0^{\pm},\\
f_2^{\pm}(V)&=-(\gamma-\hat{H}_0^{\pm})V+\frac{1}{2}\hat{H}_1^{\pm}V^2+\frac{1}{3}\hat{H}_2^{\pm}V^3+\frac{1}{4}\hat{H}_3^{\pm}V^4+M_1^{\pm},\\&\hspace{18em} V_0^{\pm}\leq V\leq V_1^{\pm},\\
f_3^{\pm}(V)&=-\gamma
V+\frac{1}{2}V^2+\frac{1}{3}D_1V^3+\frac{1}{4}D_2V^4+M_2^{\pm},\ \
V_1^{\pm}<V\leq V^*.
\end{aligned}
\right. \eqno(5.16)
$$
$g_1^{\pm}$ and $g_2^{\pm}$ are the two real roots of the equation
$$
C^{\pm}+f^{\pm}(V)=0. \ \ \eqno(5.17)
$$
Notice that for non-trivial solutions to exist, equation (5.17) must
have four real roots $\alpha_1^{\pm}\leq g_1^{\pm}\leq g_2^{\pm}\leq
\alpha_2^{\pm}$. The constants $M_1^{\pm}$ and $M_2^{\pm}$ given in
(5.16) are determined by $M_1^+=\tilde{H}_0^+$,
$M_2^+=\tilde{H}_0^++\tilde{H}_1^+V_1^++\tilde{H}_2^+{V_1^+}^2+\tilde{H}_3^+{V_1^+}^3+\tilde{H}_4^+{V_1^+}^4$,
$M_1^-=\tilde{H}_0^-+M_2^-$ and
$M_2^-=-(\tilde{H}_0^-+\tilde{H}_1^-V_0^-+\tilde{H}_2^-{V_0^-}^2+\tilde{H}_3^-{V_0^-}^3+\tilde{H}_4^-{V_0^-}^4)$
such that $f_1^{\pm}(V_0^{\pm})=f_2^{\pm}(V_0^{\pm})$ and
$f_2^{\pm}(V_1^{\pm})=f_3^{\pm}(V_1^{\pm})$, which imply that $V_X$
is continuous at $V_0^{\pm}$ and $V_1^{\pm}$.

The constant $C^{\pm}$ is determined by the following equation
$$
\frac{1}{n}=a\int_{g_1^{\pm}}^{g_2^{\pm}}\sqrt{\frac{\frac{1}{6}-D_3\tau}{C^{\pm}+f^{\pm}(\tau)}}d\tau,\
\ \ n=1, 2, 3, \cdots. \eqno(5.18)
$$
Once $C^{\pm}$ is known, from the above relationship and through
some calculations, the corresponding non-trivial solution can be
obtained from (5.15). In this paper, we only consider the
non-trivial solutions $v^{\pm}_{n1}$ and $v^{\pm}_{n2}$
corresponding to $n=1$ and $n=2$ (it can be shown that the
non-trivial solutions corresponding to large $n$ cannot be the
preferred solutions).

We choose the half thickness-length ratio $a=0.00866$. For the
purely loading and purely unloading processes, we plot the
stress-strain ($\gamma-\Delta$) curves corresponding to the constant
solutions and the first non-trivial solutions in figure 1 and 2,
respectively.
\begin{figure}
  \centering \includegraphics[width=100mm,height=75mm]{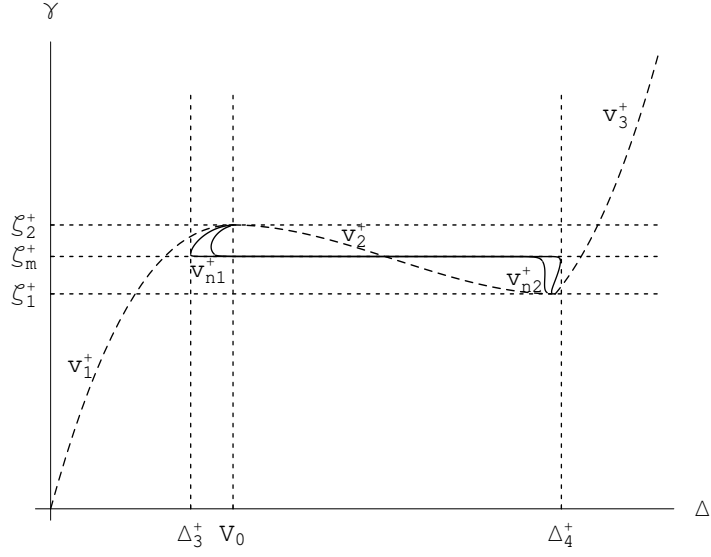}
  \caption{The $\gamma-\Delta$ curves for the constant solutions and the first two non-trivial solutions in the loading process.}
\end{figure}
\begin{figure}
  \centering \includegraphics[width=90mm,height=65mm]{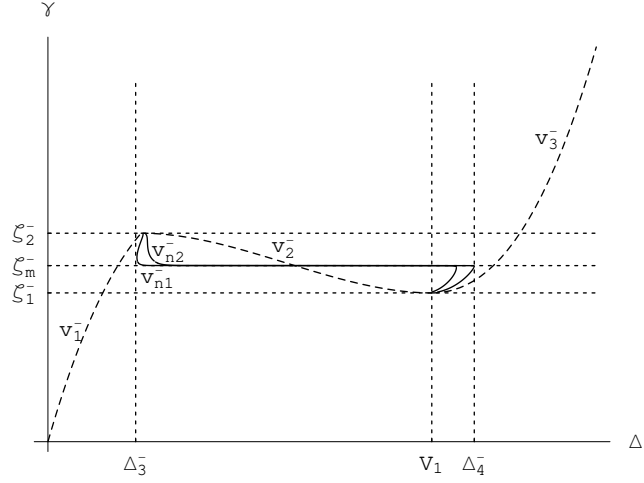}
  \caption{The $\gamma-\Delta$ curves for the constant solutions and the first two non-trivial solutions in the unloading process.}
\end{figure}
In figure 1 and 2, the three dashed curves labeled $v_1^{\pm}$,
$v_2^{\pm}$ and $v_3^{\pm}$ correspond to the three constant
solutions and the two solid curves labeled $v_{n1}^{\pm}$ and
$v_{n2}^{\pm}$ correspond to the two non-trivial solutions.

From figure 1 and 2, we can see that when
$\Delta_3^{\pm}\leq\Delta\leq \Delta_4^{\pm}$, there exist multiple
solutions. To determine which solution is the preferred one, we
compare the total pseudo-potential energies for all the possible
solutions. In the displacement-controlled problem, the total
pseudo-potential energies for the purely loading and purely
unloading processes should be given by
$$
\begin{aligned}
\Omega^+_d(V)=&{\cal{W}}^{+}(V)=\Phi_E(V)+\Phi_D^+(V)\\=&2 a E
\int_0^l(
\frac{1}{2}V^2+\frac{1}{3}D_1V^3+\frac{1}{4}D_2V^4-\frac{1}{6}a^2
VV_{XX}\\&-a^2D_3VV_X^2+\phi_d^+(V)) dX,
\end{aligned}
\eqno(5.19)
$$
and
$$
\begin{aligned}
\Omega_d^-=&{\cal{W}}^{-}(V)=\Phi_E(V)-\Phi_E^*+\Phi_D^-(V)\\
=&2 a E \int_0^l(
\frac{1}{2}V^2+\frac{1}{3}D_1V^3+\frac{1}{4}D_2V^4-\frac{1}{6}a^2
VV_{XX}\\&-a^2D_3VV_X^2+\phi_d^-(V)) dX-\Phi_E^{*}. \end{aligned}
\eqno(5.20)
$$
We plot the differences of the pseudo-potential energy
$\Delta\Omega^{\pm}_d$ between the first two non-trivial solutions
and the constant solutions in figure 3 and 4.
\begin{figure}
  \centering \includegraphics[width=100mm,height=60mm]{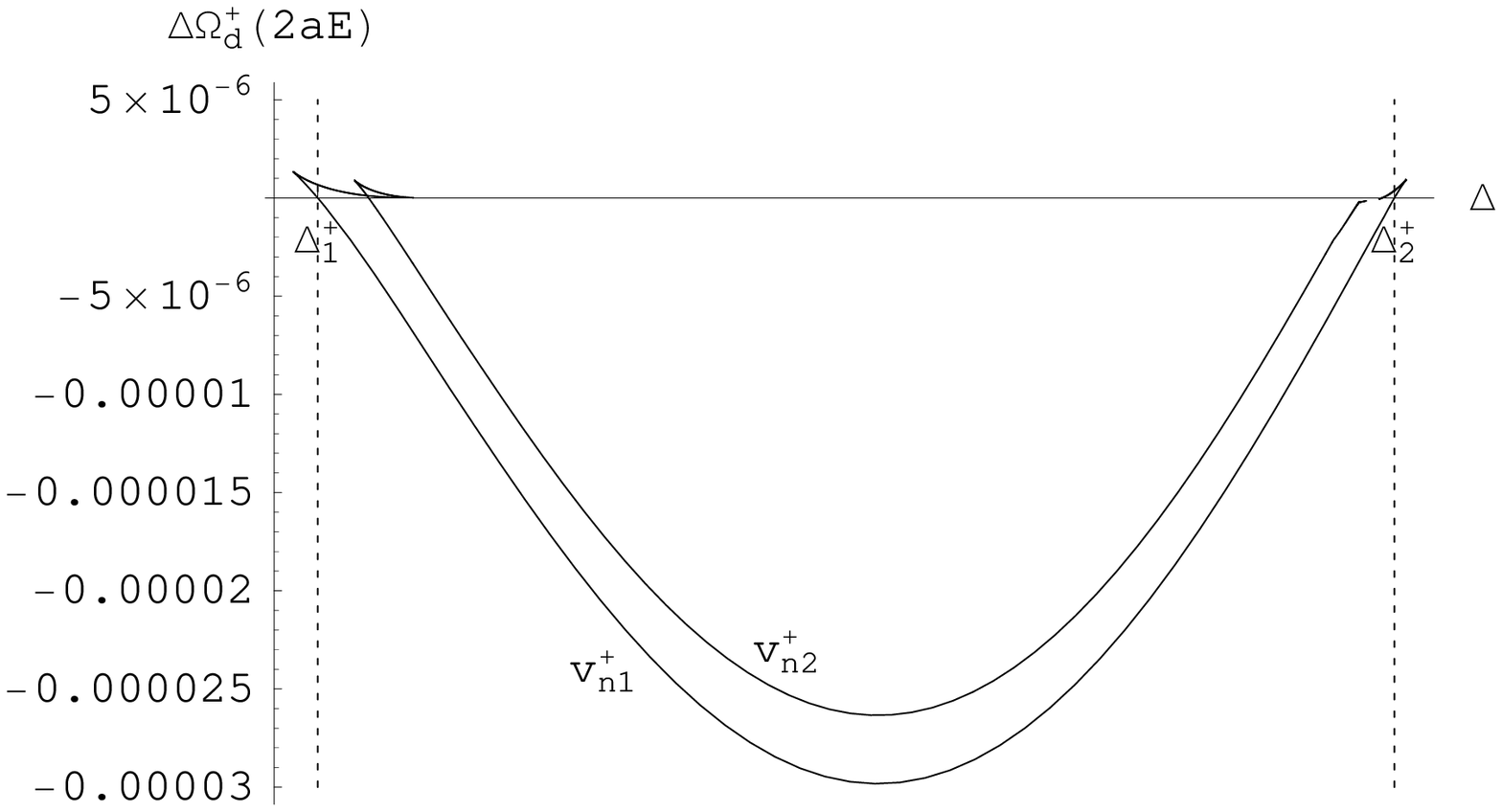}
  \caption{Differences of the pseudo-potential energy between the non-trivial solutions and the constant solutions for the loading process.}
\end{figure}
\begin{figure}
  \centering \includegraphics[width=100mm,height=60mm]{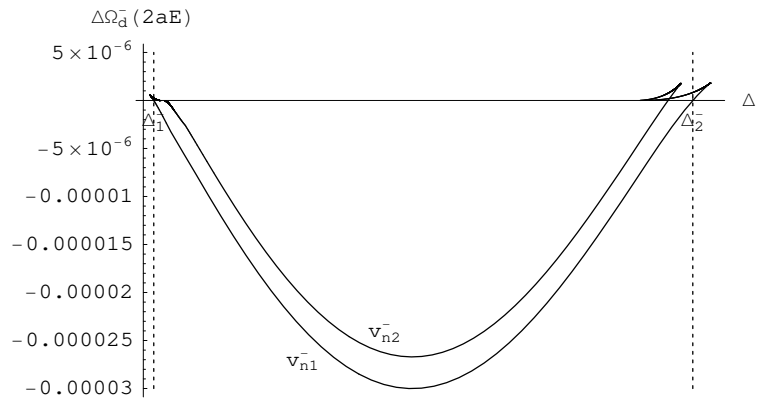}
  \caption{Differences of the pseudo-potential energy between the non-trivial solutions and the constant solutions for the unloading process.}
\end{figure}
From figure 3 and 4, we can see that for $\Delta<\Delta_1^{\pm}$ or
$\Delta>\Delta_2^{\pm}$, the constant solution is the preferred
solution and for $\Delta_1^{\pm}<\Delta<\Delta_2^{\pm}$ the first
non-trivial solution $v_{n1}^{\pm}$ is the preferred one.

In figure 5, we show the $\gamma-\Delta$ curves corresponding to the
preferred solutions, where the curve labeled ``$+$" represents the
loading part and the curve labeled ``$-$" represents the unloading
part.
\begin{figure}
  \centering \includegraphics[width=80mm,height=60mm]{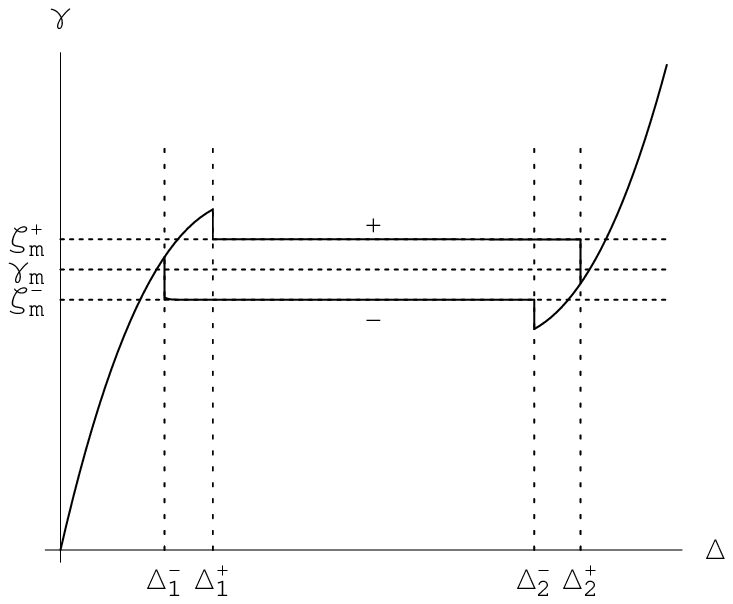}
  \caption{ $\gamma-\Delta$ curves for the preferred solutions.}
\end{figure}
We can see that the curves shown in figure 5 capture some important
features of the engineering $\gamma-\Delta$ curves measured in
experiments (see Shaw $\&$ Kyriakides 1995, 1997; Sun et al. 2000;
Tse $\&$ Sun 2000). For example, the stress peak (for the loading
part) and stress valley (for the unloading part), the stress
plateaus, the hysteresis loop and so on.

To draw figure 5, we just use the principle of minimizing the total
pseudo-potential energy. We suppose that the configuration of the
layer can jump from one metastable solution to another
metastable/stable solution once the pseudo-potential energy of the
latter one becomes smaller than the former one. But in fact, there
may exist potential barriers between two metastable solutions. In
this case, if we don't give the layer enough perturbation, the layer
will keep in its original metastable solution until this solution
become unstable or some other limitation conditions happen. Thus,
the curves shown in figure 5 are not necessarily the actual curves
measured in the experiments, as metastable solutions may be
maintained.

Alternatively, we can use the ``limit-point" instability criterion
for the onset of the nucleation or coalescence process. For the
loading part, we assume that the layer keeps in the first constant
solution until the total displacement $\Delta$ reaches $V_0$ (a
``limit-point"; see figure 1), which corresponds to the peak stress
value. As the total displacement goes on increasing, the martensite
phase will nucleate even for the homogeneous deformation, thus the
configuration of the layer jumps from the constant solution to the
first non-trivial solution, i.e., the nucleation process happens. As
$\Delta$ further increases, the deformation follows the first
non-trivial solution. Once $\Delta$ reaches $\Delta_4^+$, another
``limit-point" is also reached (see figure 1), the deformation of
the layer jumps to the third constant solution. Thus, it is at the
displacement value $\Delta=\Delta_4^+$ that coalescence process
happens. Similarly, for the unloading part, we assume that the
nucleation process happens at $\Delta=V_1$ (a ``limit-point") and
the coalescence process happens at $\Delta=\Delta_3^-$ (another
``limit-point"; see figure 2). The corresponding $\gamma-\Delta$
curves are shown in figure 6.
\begin{figure}
  \centering \includegraphics[width=80mm,height=60mm]{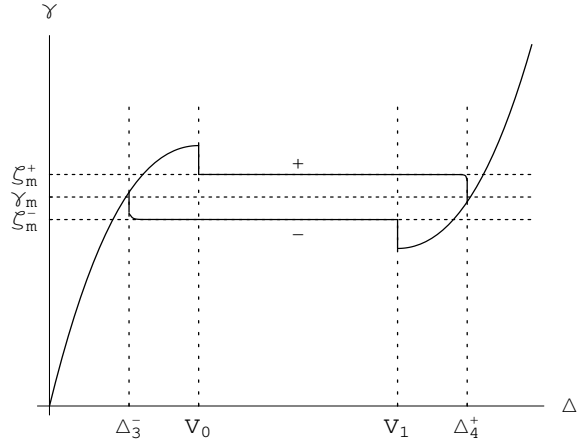}
  \caption{$\gamma-\Delta$ curves based on the ``limit-point'' criterion.}
\end{figure}
We can see that the curves shown in figure 6 are consistent with the
experimental results (see Shaw $\&$ Kyriakides 1995, 1997; Sun et
al. 2000; Tse $\&$ Sun 2000).

\section{Analysis on the coalescence process}

In this section, we shall give some analysis on the coalescence
process, which plays a central role in the whole phase transition
process. The following analysis of this section are based on the
analytical solutions obtained in the section 5. Here we wish to give
some descriptions and explanations for the origin of the instability
during the coalescence process, the accompanying stress drop/jump
and the morphology varies of the specimen. We shall also consider
the size-effect of the specimens on the coalescence process.

As mentioned before, the coalescence process of phase fronts
observed in quasi-static experiments is a dynamic one, which is an
indication that an instability occurs. In subsection 5.2, we have
already used the minimum energy principle to model the coalescence
process (see figure 5). By comparing the total pseudo-potential
energies of the constant solutions and nontrivial solutions, we
found that the coalescence process should take place at the point
$\Delta=\Delta_2^+$ for the loading case and $\Delta=\Delta_1^-$ for
the unloading case. Besides the minimum energy principle, another
possible reason for the onset of the coalescence process could be
the fact that some ``limit-points'' have been reached. In subsection
5.2, we also used the ``limit-points'' instability criterion to
model the coalescence process (see figure 6). We found that it was
at the points $\Delta=\Delta_4^+$ for the loading case and
$\Delta=\Delta_3^-$ for the unloading case that some
``limit-points'' had been reached and the coalescence process took
place.

\noindent\textbf{Remark}: In the sequel, we shall only use the
``limit-points'' instability criterion to model the coalescence
process.

Systematic experimental results have obviously shown that the
coalescence process is inevitably accompanied the varies of the
stress value and the surface morphology of the specimen. Here we
shall use our model to describe these phenomena. For the purpose of
clearness, we consider the second nontrivial solution (i.e., we
choose $n=2$ in equation (5.18)) and consider the case that the
phase fronts coalesce at the middle part of the layer. Thus the
transformation scheme can be simply written as $MA+AM\rightarrow M$
for the loading case and $AM+MA\rightarrow A$ for the unloading
case. Figure 7(a) shows the engineering stress-strain curve
corresponding to the coalescence process during the loading case
($a=0.0043$). Corresponding to the 5 points $A$--$E$ show in figure
7(a), we plot the profiles of the layer in figure 7(b).
\begin{figure}
   \centering
  \begin{minipage}{0.45\textwidth}
     \centering \includegraphics[width=65mm,height=48mm]{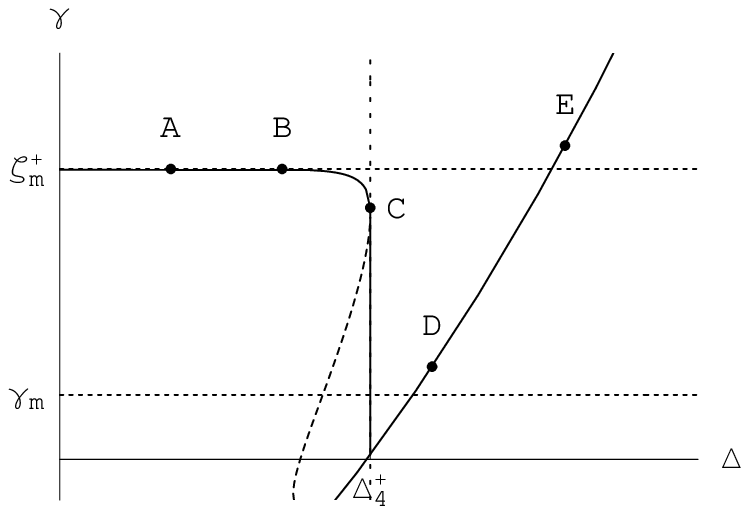}\\(a)
  \end{minipage}
  \begin{minipage}{0.45\textwidth}
     \centering \includegraphics[width=60mm,height=42mm]{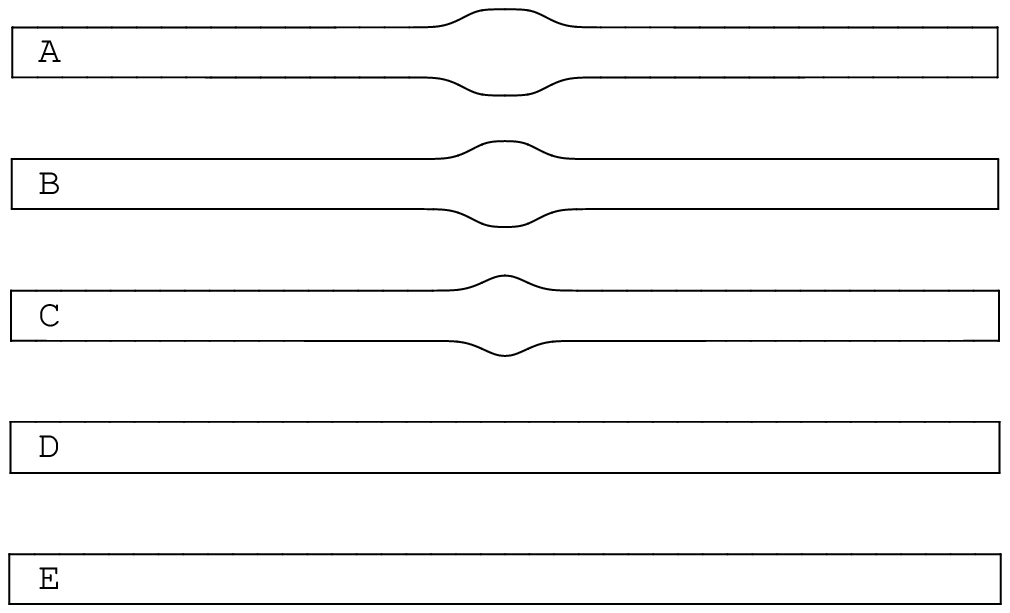}\\(b)
  \end{minipage}
  \caption{(a) Engineering stress-strain curve for the
coalescence process during the loading case ($a=0.0043$). (b)
Profiles of the layer corresponding to the 5 points $A$--$E$.}
\end{figure}
Here the radial deformation has been enlarged for clearness. From
figure 7, we can see that the stress drop and the deformation
process of the layer are very similar to the actual experimental
procedures. Figure 8 shows the stress jump and the deformation
process of the layer corresponding to the coalescence process during
the unloading case ($a=0.0043$).
\begin{figure}
   \centering
  \begin{minipage}{0.45\textwidth}
     \centering \includegraphics[width=65mm,height=48mm]{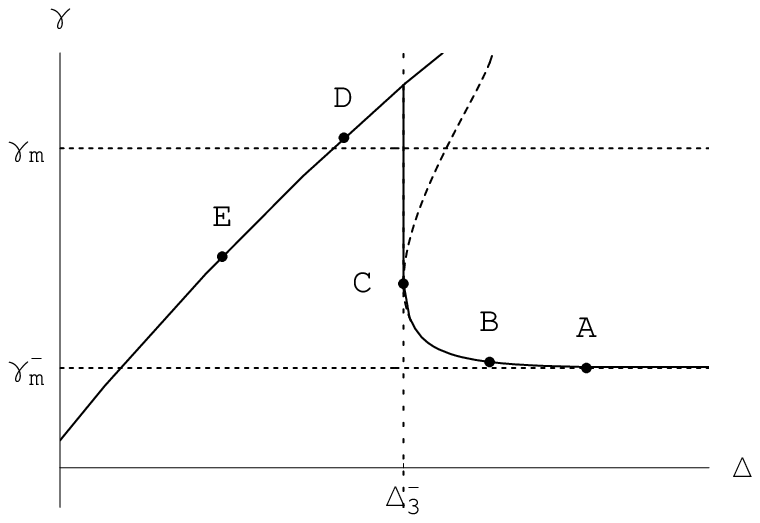}\\(a)
  \end{minipage}
  \begin{minipage}{0.45\textwidth}
     \centering \includegraphics[width=60mm,height=42mm]{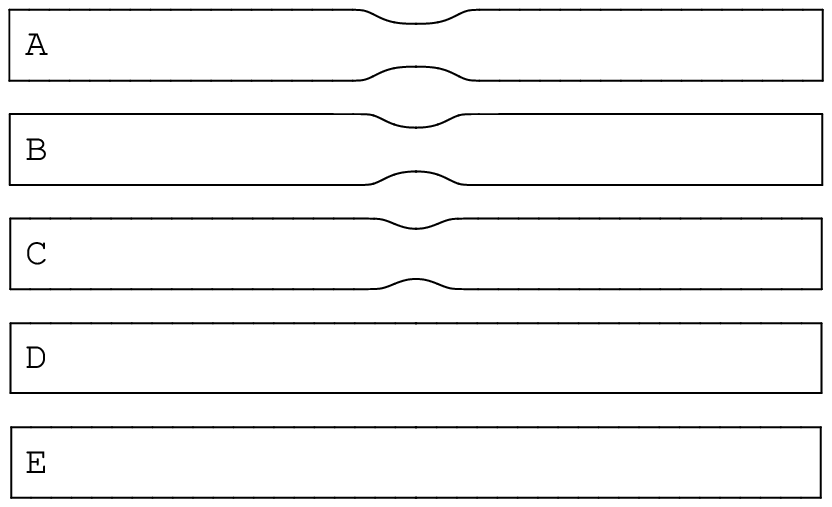}\\(b)
  \end{minipage}
  \caption{(a) Engineering stress-strain curve for the
coalescence process during the unloading case ($a=0.0043$). (b)
Profiles of the layer corresponding to the 5 points $A$--$E$.}
\end{figure}
We can see that the coalescence process shown in figure 8 is also
consistent with the experimental results.

Next, we consider the influence of the geometric size effect of the
layer on the coalescence process. From (5.15) and (5.18), we can see
the nontrivial solutions obtained here depend directly on the half
thickness-length ratio of the layer, which shows the fact that our
model can reflect some important information on the geometric size
effect. Here we only consider the loading case, and the similar
result can also be derived for the unloading case. Figure 9 shows
the $\gamma-\Delta$ curves of the second nontrivial solutions
($n=2$) corresponding to some different half thickness-length ratio.
\begin{figure}
  \centering \includegraphics[width=80mm,height=60mm]{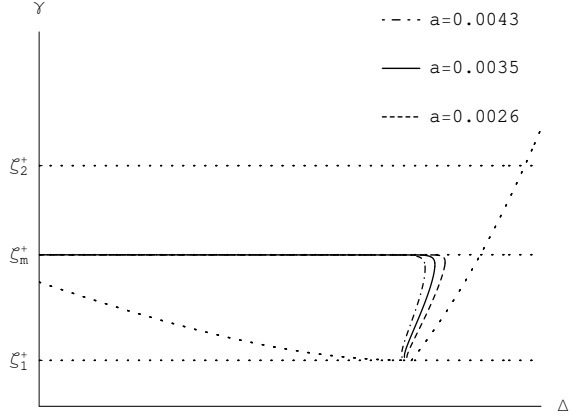}
  \caption{$\gamma-\Delta$ curves of the second nontrivial solutions corresponding to some different half thickness-length ratio ($a=0.0026$, $0.0035$, $0.0043$).}
\end{figure}
From figure 9, we can see that as $a$ decreases, the $\gamma-\Delta$
curve of the non-trivial solution moves towards the constant
solution. Actually, one can prove that the $\gamma-\Delta$ curve of
the non-trivial solution moves toward first constant solution when
$\xi_m^+<\gamma<\xi_2^+$ and towards the third constant solution
when $\xi_1^+<\gamma<\xi_m^+$. The following is the proof for the
case of the second nontrivial solutions ($n=2$) and
$\xi_1^+<\gamma<\xi_m^+$ (the case of other nontrivial solutions and
$\xi_m^+<\gamma<\xi_2^+$ can be proved similarly).

First, for the second nontrivial solution, we have the following
equation (cf. (5.18))
$$
\frac{1}{2}=a \int_{g_1^+}^{g_2^+}\sqrt{\frac{\frac{1}{6}-D_3
V}{C^++f^+(V)}}dV,\ \ \ \eqno(6.1)
$$
which can be considered as a relationship between the half
thickness-length ratio $a$ and the constant $C^+$. Through some
simple analysis, we find that the equation $C^++f^+(V)=0$ has four
real roots $\alpha_1^+\leq g_1^+\leq g_2^+\leq \alpha_2^+$ if and
only if
$$
C_2^+\leq C^+ \leq \textrm{min}(C_1^+,C_3^+),\ \ \eqno(6.2)
$$
where
$$
C_i^+=-f_i^+(v_i^+),\ \ \ i=1,2,3.
$$
In this case, the function $C^++f^+(V)$ can be written in the
following form
$$
C^++f^+(V)=f_C(V)(V-\alpha_1^+)(V-g_1^+)(V-g_2^+)(V-\alpha_2^+), \ \
\eqno(6.3)
$$
where $f_C(V)$ is a non-zero and bounded continuous function
depending on the constant $C$.

Through some further analysis, we find that when
$\xi_1^+<\gamma<\xi_m^+$, $C^+$ should satisfy $C_2^+\leq C^+\leq
C_3^+$ and
$$
\lim_{C^+\rightarrow C_3^+}\alpha_1^+<v_1^+<\lim_{C^+\rightarrow
C_3^+}g_1^+<v_2^+<\lim_{C^+\rightarrow
C_3^+}g_2^+=v_3^+=\lim_{C^+\rightarrow C_3^+}\alpha_2^+,\ \
\eqno(6.4)
$$
where $v_1^+$, $v_2^+$ and $v_3^+$ are the three real roots of
equation (5.14). From (6.1) and (6.3), we have
$$
a=\frac{1}{2 \int_{g_1^+}^{g_2^+}\sqrt{\frac{1/6-D_3
V}{f_C(V)(V-\alpha_1^+)(V-g_1^+)(V-g_2^+)(V-\alpha_2^+)}}dV}.\ \
\eqno(6.5)
$$
Thus as $a$ tends to zero, the integration
$\int_{g_1^+}^{g_2^+}\sqrt{\frac{1/6-D_3
V}{f_C(V)(V-\alpha_1^+)(V-g_1^+)(V-g_2^+)(V-\alpha_2^+)}}dV$ should
tends to infinity, and this is equivalent to $C^+$ tending to
$C_3^+$. By using (5.15) and (6.5), we can give the following
derivation
$$
\begin{aligned}
\Delta&=\int_0^1VdZ=2\int_0^{\frac{1}{2}}VdZ=2\int_{g_1^+}^{g_2^+}V\frac{dZ}{dV}dV\\
&=2a \int_{g_1^+}^{g_2^+}V\sqrt{\frac{1/6-D_3
V}{f_C(V)(V-\alpha_1^+)(V-g_1^+)(V-g_2^+)(V-\alpha_2^+)}}dV\\
&=\frac{\int_{g_1^+}^{g_2^+}V\sqrt{\frac{1/6-D_3
V}{f_C(V)(V-\alpha_1^+)(V-g_1^+)(V-g_2^+)(V-\alpha_2^+)}}dV}{\int_{g_1^+}^{g_2^+}\sqrt{\frac{1/6-D_3
V}{f_C(V)(V-\alpha_1^+)(V-g_1^+)(V-g_2^+)(V-\alpha_2^+)}}dV}\\
&=g_2^+-\frac{\int_{g_1^+}^{g_2^+}(g_2^+-V)\sqrt{\frac{1/6-D_3
V}{f_C(V)(V-\alpha_1^+)(V-g_1^+)(V-g_2^+)(V-\alpha_2^+)}}dV}{\int_{g_1^+}^{g_2^+}\sqrt{\frac{1/6-D_3
V}{f_C(V)(V-\alpha_1^+)(V-g_1^+)(V-g_2^+)(V-\alpha_2^+)}}dV}\\
&=g_2^+-\frac{\int_{g_1^+}^{g_2^+}\sqrt{\frac{(1/6-D_3
V)(g_2^+-V)}{f_C(V)(V-\alpha_1^+)(V-g_1^+)(\alpha_2^+-V)}}dV}{\int_{g_1^+}^{g_2^+}\sqrt{\frac{1/6-D_3
V}{f_C(V)(V-\alpha_1^+)(V-g_1^+)(V-g_2^+)(V-\alpha_2^+)}}dV}.
\end{aligned}
\eqno(6.6)
$$
By using (6.4), it is easy to prove
$$
\lim_{C\rightarrow C_3^+}\int_{g_1^+}^{g_2^+}\sqrt{\frac{(1/6-D_3
V)(g_2^+-V)}{f_C(V)(V-\alpha_1^+)(V-g_1^+)(\alpha_2^+-V)}}dV<+\infty
$$
and
$$
\lim_{C\rightarrow C_3^+}\int_{g_1^+}^{g_2^+}\sqrt{\frac{1/6-D_3
V}{f_C(V)(V-\alpha_1^+)(V-g_1^+)(V-g_2^+)(V-\alpha_2^+)}}dV=+\infty.
$$
Thus, as $a$ tends to zero, the total elongation of the layer should
satisfy
$$
\lim_{a\rightarrow 0}\Delta=\lim_{C\rightarrow C_3^+}\Delta=\lim_{C
\rightarrow C_3^+}g_2^+=v_3^+.\ \ \ \eqno(6.7)
$$
This means that the $\gamma - \Delta$ curve of the second nontrivial
solution tends to that of the third constant solution as $a$ tends
to zero when $\xi_1^+<\gamma<\xi_m^+$.

Figure 9 shows that the $\gamma - \Delta$ curve of the nontrivial
solution has a snap-back structure. But in a purely loading process,
the total elongation $\Delta$ cannot decrease. Thus the state of the
layer should jump from the nontrivial solution to the constant
solution at some special point, which corresponds to the coalescence
process. Here we also use the ``limit-points'' instability criterion
to model the coalescence process. The actual $\gamma - \Delta$
curves of the coalescence processes corresponding to some different
half thickness-length ratio are shown in figure 10.
\begin{figure}
  \centering \includegraphics[width=80mm,height=60mm]{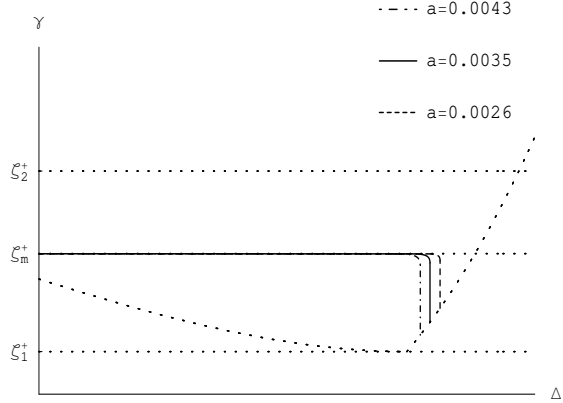}
  \caption{Engineering stress-strain curves for the coalescence process during the loading case ($a=0.0026$, $0.0035$, $0.0043$).}
\end{figure}
From figure 10, we can see that as $a$ decreases, the stress drop
during the coalescence process also decreases. Actually, (6.7)
implies that the limit point should tend to the third constant
solution as $a$ tends to zero. Thus, the stress drop should tend to
zero as $a$ tends to zero. Similarly, in the process of
AM+MA$\rightarrow$A during the unloading case, the stress jump
should tend to zero as $a$ tends to zero. Thus we can say that the
geometrical size of the specimen plays an important role in the
coalescence process.


\section{Conclusions}

In this paper, we aim to study the phase fronts coalescence
phenomena during the phase transition process in a thin SMA layer.
For that purpose, we derived a quasi-2D model with a non-convex
effective strain energy function while taking into account the
rate-independent dissipation effect.

In a two-dimensional setting, we studied the symmetric deformation
of a thin SMA layer. Based on the field equations and the
traction-free boundary conditions, by using the coupled
series-asymptotic expansion method, we expressed the total elastic
potential energy of the layer $\Phi_E$ as a function of the leading
order term of the axial strain $V(X)$. We further considered the
mechanical dissipation effect in a purely one-dimensional setting
and expressed the total amount of mechanical dissipations $\Phi_D^+$
(loading case) and $\Phi_D^-$ (unloading case) as functions of the
axial strain $V(X)$. The equilibrium equation was then determined by
using the principle of maximizing the total energy dissipation. We
further considered an illustrative example with some given material
constants and some special form of dissipation density functions.
With the free end boundary conditions, we managed to construct the
analytical solutions for both a force-controlled problem and a
displacement-controlled problem.

Based on the analytical solutions obtained and by using the
limit-point instability criterion, we studied the phase fronts
coalescence phenomena. It was revealed that during the coalescence
process, the configurations of the layer switched from the
nontrivial solution modes to the trivial solution modes, which was
caused by the presence of the ``limit points". The morphology varies
of the layer and the accompanying stress drop/jump during the
coalescence process can be described. The influence of the
thickness-length ratio of the specimen on the coalescence process
was also studied. It was found that the zero limit of the
thickness-length ratio can lead to the smooth switch of nontrivial
modes to trivial modes with no stress drop or stress jump.

\makeatletter
\def\app@number#1{ \setcounter{#1}{0}%
\@addtoreset{#1}{section}%
\@namedef{the#1}{\thesection.\arabic{#1}}}
\def\appendix{\@ifstar{\appendix@star}{\appendix@nostar}}
\def\appendix@nostar{%
\def\lb@section{ \appendixname \ \thesection.\half@em}
\def\lb@empty@section{\appendixname\ \thesection}
\setcounter{section}{0}\def\thesection{Appendix \Alph{section}}%
\setcounter{subsection}{0}%
\setcounter{subsubsection}{0}%
\setcounter{paragraph}{0}%
\app@number{equation}\app@number{figure}\app@number{table}}
\def\appendix@star{%
\def\lb@section{\appendixname}\let\lb@empty@section\lb@section
\setcounter{section}{0}\def\thesection{Appendix \Alph{section}}%
\setcounter{subsection}{0}%
\setcounter{subsubsection}{0}%
\setcounter{paragraph}{0}%
\app@number{equation}\app@number{figure}\app@number{table}}
\makeatother

\appendix

\section{Incremental elastic moduli}

For initially isotropic material, in the case that there are no
prestresses, the elastic potential energy function $\Psi$ should be
a function of the principal stretches $\lambda_1$ and $\lambda_2$,
namely $\Psi$=$\Psi(\lambda_1,\lambda_2)$. Denote by
$\Psi_j=\frac{\partial \Psi}{\partial
\lambda_j}|_{\lambda_1=\lambda_2=1}$, then $\Psi_1=\Psi_2$ should
vanish since there are no prestresses.

The non-zero first order incremental elastic moduli can be written
as
$$
\begin{aligned}
&\tau_1=a^1_{1111}=\Psi_{11},\\
&\tau_2=a^1_{1122}=\Psi_{12},\\
&\tau_3=a^1_{1212}=\frac{1}{2}(\tau_1-\tau_2),\\
&\tau_4=a^1_{1221}=\tau_3.
\end{aligned}
$$
There are only two independent constants among $\tau_i$.

The non-zero second order incremental elastic moduli can be written
as
$$
\begin{aligned}
&\eta_1=a^2_{111111}=\Psi_{111},\\
&\eta_2=a^2_{111122}=\Psi_{112},\\
&\eta_3=a^2_{111212}=\frac{1}{4}(\tau_1+\tau_2+\eta_1-\eta_2),\\
&\eta_4=a^2_{111221}=\frac{1}{4}(-\tau_1-\tau_2+\eta_1-\eta_2),\\
\end{aligned}
$$
There are only two additional independent constants among $\eta_i$.

The non-zero third order incremental elastic moduli can be written
as
$$
\begin{aligned}
&\theta_1=a^3_{11111111}=\Psi_{1111},\\
&\theta_2=a^3_{11111122}=\Psi_{1112},\\
&\theta_3=a^3_{11111212}=\frac{1}{12}(-3\tau_1-3\tau_2+3\eta_1+3\eta_2+2\theta_1-2\theta_2),\\
&\theta_4=a^3_{11111221}=\frac{1}{12}(3\tau_1+3\tau_2-3\eta_1-3\eta_2+2\theta_1-2\theta_2),\\
&\theta_5=a^3_{11112222}=\Psi_{1122},\\
&\theta_6=a^3_{11121222}=\frac{1}{12}(-3\tau_1-3\tau_2+6\eta_2+\theta_1+2\theta_2-3\theta_5),\\
&\theta_7=a^3_{11122122}=\frac{1}{12}(3\tau_1+3\tau_2-6\eta_2+\theta_1+2\theta_2-3\theta_5),\\
&\theta_{8}=a^3_{12121212}=\frac{1}{8}(3\tau_1+3\tau_2+6\eta_1-6\eta_2+\theta_1-4\theta_2+3\theta_5),\\
&\theta_{9}=a^3_{12121221}=\frac{1}{8}(-3\tau_1-3\tau_2+\theta_1-4\theta_2+3\theta_5),\\
&\theta_{10}=a^3_{12122121}=\frac{1}{8}(3\tau_1+3\tau_2-2\eta_1+2\eta_2+\theta_1-4\theta_2+3\theta_5),\\
\end{aligned}
$$
There are only three additional independent constants among
$\theta_i$.

\section{Non-dimensional field equations}

The full forms of the non-dimensional field equations (3.15) and
(3.16) are given by:
$$
\begin{aligned}
&{2 {\tau_3} u_s+({\tau_2}+{\tau_3}) w_x+s (4 {\tau_3} u_{ss}+(2
{\tau_2}+2 {\tau_3}) w_{xs})+{\tau_1} u_{xx}+\epsilon (2 {\eta_3} w
u_s+2 {\eta_3} u_s u_x}\\& {+({\eta_2}+{\eta_4}) w
w_x+({\eta_2}+{\eta_4}) u_x w_x+s^2 (8 {\eta_3} w_s u_{ss}+8
{\eta_3} u_s w_{ss}+4 {\eta_4} w_{ss} w_x} \\&{+(4 {\eta_2}+4
{\eta_4}) w_s w_{xs})+{\eta_2} w u_{xx}+{\eta_1} u_x u_{xx}+s (16
{\eta_3} u_s w_s+4 {\eta_3} w u_{ss}+4 {\eta_3} u_{ss} u_x}\\& {+(2
{\eta_2}+8 {\eta_4}) w_s w_x+8 {\eta_3} u_s u_{xs}+4 {\eta_4} w_x
u_{xs}+(2 {\eta_2}+2 {\eta_4}) w w_{xs}+(2 {\eta_2}+2 {\eta_4}) u_x
w_{xs}}\\& {+2 {\eta_2} w_s u_{xx}+2 {\eta_4} u_s w_{xx}+{\eta_3}
w_x w_{xx}))+\epsilon^2(\theta_3 w^2 u_s+2\theta_6 w u_s
u_x+\theta_3 u_s
u_x^2}\\
&+(\frac{\theta_2}{2}+\frac{\theta_4}{2}) w^2 w_x+(\theta_5+\theta_
7) w u_x w_x+(\frac{\theta_2}{2}+\frac{\theta_4}{2}) u_x^2 w_x+ s^3
(8 \theta_3 w_s^2 u_{ss}\\& +16 \theta_3 u_s w_s w_{ss}+8 \theta_4
w_s w_{ss} w_x+(4 \theta_2+4 \theta_4) w_s^2 w_{xs})+\frac{1}{2}
\theta_5 w^2 u_{xx}+\theta_2 w u_x u_{xx}\\& +\frac{1}{2} \theta_1
u_x^2 u_{xx}+ s^2 (28 \theta_3 u_s w_s^2+8 \theta_8 u_s^2 u_{ss}+8
\theta_3 w w_s u_{ss}+8 \theta_3 w u_s w_{ss}+8 \theta_6 w_s u_{ss}
u_x\\&+8 \theta_6 u_s w_{ss} u_x+(2 \theta_2+14 \theta_4) w_s^2
w_x+8 \theta_9 u_s u_{ss} w_x+4 \theta_4 w w_{ss} w_x+4 \theta_7
w_{ss} u_x w_x\\& +2 \theta_{10} u_{ss} w_x^2+16 \theta_6 u_s w_s
u_{xs}+8 \theta_7 w_s w_x u_{xs}+(4 \theta_6+4 \theta_9) u_s^2
w_{xs}+(4 \theta_2+4 \theta_4) w w_s w_{xs}\\& +(4 \theta_5+4
\theta_7) w_s u_x w_{xs}+(4 \theta_{10}+4 \theta_7) u_s w_x
w_{xs}+(\theta_6+\theta_9) w_x^2 w_{xs}+2 \theta_5 w_s^2 u_{xx}\\&
+4 \theta_7 u_s w_s w_{xx}+2 \theta_6 w_s w_x w_{xx})+ s (4 \theta_8
u_s^3+16 \theta_3 w u_s w_s+2 \theta_3 w^2 u_{ss}+16 \theta_6 u_s
w_s u_x\\& +4 \theta_6 w u_{ss} u_x+2 \theta_3 u_{ss} u_x^2+(2
\theta_6+6 \theta_9) u_s^2 w_x+(2 \theta_2+8 \theta_4) w w_s w_x+(3
\theta_{10}+2 \theta_7) u_s w_x^2\\&+(2 \theta_5+8 \theta_7) w_s u_x
w_x+(\frac{\theta_6}{2}+\frac{\theta_9}{2}) w_x^3+8 \theta_6 w u_s
u_{xs}+8 \theta_3 u_s u_x u_{xs}+4 \theta_7 w w_x u_{xs}\\&+4
\theta_4 u_x w_x u_{xs}+(\theta_2+\theta_4) w^2 w_{xs}+(2 \theta_5+2
\theta_7) w u_x w_{xs}+(\theta_2+\theta_4) u_x^2 w_{xs}+2 \theta_3
u_s^2 u_{xx}\\& +2 \theta_5 w w_s u_{xx}+2 \theta_2 w_s u_x u_{xx}+2
\theta_4 u_s w_x u_{xx}+\frac{1}{2} \theta_3 w_x^2 u_{xx}+2 \theta_7
w u_s w_{xx}+2 \theta_4 u_s u_x w_{xx}\\&+ \theta_6 w w_x
w_{xx}+\theta_3 u_x w_x w_{xx}))=0,
\end{aligned}
\eqno(B1)
$$

$$
\begin{aligned}
& 6 \tau_1 w_s+(2\tau_2+2\tau_3) u_{xs}+\tau_3w_{xx}+4 s\tau_1
w_{ss}+\epsilon ({4 {\eta_3} u_s^2+6 {\eta_1} w w_{s}}\\& {+8 s^2
{\eta_1} w_{s} w_{ss}+6 {\eta_2} w_{s} u_x+6 {\eta_4} u_s w_{x}+2
{\eta_3} w_{x}^2+(2 {\eta_2}+2 {\eta_4}) w u_{xs}}\\&{+(2 {\eta_2}+2
{\eta_4}) u_x u_{xs}+2 {\eta_4} u_s u_{xx}+{\eta_3} w_{x}
u_{xx}+{\eta_3} w w_{xx}+{\eta_3} u_x w_{xx}}\\& {+s (12 {\eta_1}
w_{s}^2+8 {\eta_3} u_s u_{ss}+4 {\eta_1} w w_{ss}+4 {\eta_2} w_{ss}
u_x+4 {\eta_4} u_{ss} w_{x}}\\& {+(4 {\eta_2}+4 {\eta_4}) w_{s}
u_{xs}+8 {\eta_4} u_s w_{xs}+4 {\eta_3} w_{x} w_{xs}+2 {\eta_3}
w_{s} w_{xx})})\\&+\epsilon^2({4 {\theta_3} w u_s^2+3 {\theta_1} w^2
w_{s}+8 s^3 {\theta_1} w_{s}^2 w_{ss}+4 {\theta_6} u_s^2 u_x+6
{\theta_2} w w_{s} u_x+3 {\theta_5} w_{s} u_x^2}\\& {+6 {\theta_4} w
u_s w_{x}+6 {\theta_7} u_s u_x w_{x}+2 {\theta_3} w w_{x}^2+2
{\theta_6} u_x w_{x}^2+({\theta_2}+{\theta_4}) w^2 u_{xs}}\\& {+(2
{\theta_5}+2 {\theta_7}) w u_x u_{xs}+({\theta_2}+{\theta_4}) u_x^2
u_{xs}+2 {\theta_7} w u_s u_{xx}+2 {\theta_4} u_s u_x u_{xx}}\\&
{+{\theta_6} w w_{x} u_{xx}+{\theta_3} u_x w_{x} u_{xx}+\frac{1}{2}
{\theta_3} w^2 w_{xx}+{\theta_6} w u_x w_{xx}+\frac{1}{2} {\theta_3}
u_x^2 w_{xx}}\\& {+s^2 (12 {\theta_1} w_{s}^3+16 {\theta_3} u_s
w_{s} u_{ss}+8 {\theta_3} u_s^2 w_{ss}+8 {\theta_1} w w_{s} w_{ss}+8
{\theta_2} w_{s} w_{ss} u_x}\\& {+8 {\theta_4} w_{s} u_{ss} w_{x}+8
{\theta_4} u_s w_{ss} w_{x}+2 {\theta_3} w_{ss} w_{x}^2+(4
{\theta_2}+4 {\theta_4}) w_{s}^2 u_{xs}}\\& {+16 {\theta_4} u_s
w_{s} w_{xs}+8 {\theta_3} w_{s} w_{x} w_{xs}+2 {\theta_3} w_{s}^2
w_{xx})+s (20 {\theta_3} u_s^2 w_{s}+12 {\theta_1} w w_{s}^2}\\& {+8
{\theta_3} w u_s u_{ss}+2 {\theta_1} w^2 w_{ss}+12 {\theta_2}
w_{s}^2 u_x+8 {\theta_6} u_s u_{ss} u_x+4 {\theta_2} w w_{ss}
u_x}\\& {+2 {\theta_5} w_{ss} u_x^2+24 {\theta_4} u_s w_{s} w_{x}+4
{\theta_4} w u_{ss} w_{x}+4 {\theta_7} u_{ss} u_x w_{x}+7 {\theta_3}
w_{s} w_{x}^2}\\& {+(4 {\theta_6}+4 {\theta_9}) u_s^2 u_{xs}+(4
{\theta_2}+4 {\theta_4}) w w_{s} u_{xs}+(4 {\theta_5}+4 {\theta_7})
w_{s} u_x u_{xs}+}\\& {(4 {\theta_{10}}+4 {\theta_7}) u_s w_{x}
u_{xs}+({\theta_6}+{\theta_9}) w_{x}^2 u_{xs}+8 {\theta_4} w u_s
w_{xs}+8 {\theta_7} u_s u_x w_{xs}}\\& {+4 {\theta_3} w w_{x}
w_{xs}+4 {\theta_6} u_x w_{x} w_{xs}+4 {\theta_7} u_s w_{s} u_{xx}+2
{\theta_6} w_{s} w_{x} u_{xx}+2 {\theta_10} u_s^2 w_{xx}}\\& {+2
{\theta_3} w w_{s} w_{xx}+2 {\theta_6} w_{s} u_x w_{xx}+2 {\theta_9}
u_s w_{x} w_{xx}+\frac{1}{2} {\theta_8} w_{x}^2 w_{xx})})=0.
\end{aligned}
\eqno(B2)
$$

\section{}
The formulas of the constants $a_i$ ($i=1,\cdots,25$) in
(3.26)-(3.28) are given below:
$$
\begin{aligned}
&{ {a_ 1}=-\frac{{\eta_ 2}}{2 {\tau_ 3}}+\frac{{\tau_ 2} {\eta_
3}}{2 {{{\tau_ 3}}^2}}+\frac{{\eta_ 3}}{2 {\tau_ 3}}-\frac{{\eta_
4}}{2 {\tau_ 3}} ,}\\& { {a_2}=-\frac{{\eta_ 2}}{2 {\tau_
3}}+\frac{{\tau_ 2} {\eta_ 3}}{2 {{{\tau_ 3}}^2}}+\frac{{\eta_ 3}}{2
{\tau_ 3}}-\frac{{\eta_ 4}}{2 {\tau_ 3}} ,}\\& { {a_3}=-\frac{{\eta_
2}}{2 {\tau_ 3}}+\frac{{\tau_ 1} {\eta_ 3}}{2 {{{\tau_ 3}}^2}} ,}\\&
{ {a_4}=-\frac{{\eta_ 1}}{2 {\tau_ 3}}+\frac{{\tau_ 1} {\eta_ 3}}{2
{{{\tau_ 3}}^2}} ,}\\& { {a_5}=\frac{{\eta_ 2} {\eta_ 3}}{2 {{{\tau_
3}}^2}}-\frac{{\tau_ 2} {{{\eta_ 3}}^2}}{2 {{{\tau_
3}}^3}}-\frac{{{{\eta_ 3}}^2}}{2 {{{\tau_ 3}}^2}}+\frac{{\eta_ 3}
{\eta_ 4}}{2 {{{\tau_ 3}}^2}}-\frac{{\theta_ 2}}{4 {\tau_
3}}+\frac{{\tau_ 2} {\theta_ 3}}{4 {{{\tau_ 3}}^2}}+\frac{{\theta_
3}}{4 {\tau_ 3}}-\frac{{\theta_ 4}}{4 {\tau_ 3}} ,}\\& {
{a_6}=\frac{{\eta_ 2} {\eta_ 3}}{{{{\tau_ 3}}^2}}-\frac{{\tau_ 2}
{{{\eta_ 3}}^2}}{{{{\tau_ 3}}^3}}-\frac{{{{\eta_ 3}}^2}}{{{{\tau_
3}}^2}}+\frac{{\eta_ 3} {\eta_ 4}}{{{{\tau_ 3}}^2}}-\frac{{\theta_
5}}{2 {\tau_ 3}}+\frac{{\tau_ 2} {\theta_ 6}}{2 {{{\tau_
3}}^2}}+\frac{{\theta_ 6}}{2 {\tau_ 3}}-\frac{{\theta_ 7}}{2 {\tau_
3}} ,}\\& { {a_7}=\frac{{\eta_ 2} {\eta_ 3}}{2 {{{\tau_
3}}^2}}-\frac{{\tau_ 2} {{{\eta_ 3}}^2}}{2 {{{\tau_
3}}^3}}-\frac{{{{\eta_ 3}}^2}}{2 {{{\tau_ 3}}^2}}+\frac{{\eta_ 3}
{\eta_ 4}}{2 {{{\tau_ 3}}^2}}-\frac{{\theta_ 2}}{4 {\tau_
3}}+\frac{{\tau_ 2} {\theta_ 3}}{4 {{{\tau_ 3}}^2}}+\frac{{\theta_
3}}{4 {\tau_ 3}}-\frac{{\theta_ 4}}{4 {\tau_ 3}} ,}\\& {
{a_8}=\frac{{\eta_ 2} {\eta_ 3}}{2 {{{\tau_ 3}}^2}}-\frac{{\tau_ 1}
{{{\eta_ 3}}^2}}{2 {{{\tau_ 3}}^3}}+\frac{{\tau_ 1} {\theta_ 3}}{4
{{{\tau_ 3}}^2}}-\frac{{\theta_ 5}}{4 {\tau_ 3}} ,}\\& {
{a_9}=\frac{{\eta_ 1} {\eta_ 3}}{2 {{{\tau_ 3}}^2}}+\frac{{\eta_ 2}
{\eta_ 3}}{2 {{{\tau_ 3}}^2}}-\frac{{\tau_ 1} {{{\eta_
3}}^2}}{{{{\tau_ 3}}^3}}-\frac{{\theta_ 2}}{2 {\tau_
3}}+\frac{{\tau_ 1} {\theta_ 6}}{2 {{{\tau_ 3}}^2}} ,}\\& {
{a_{10}}=\frac{{\eta_ 1} {\eta_ 3}}{2 {{{\tau_ 3}}^2}}-\frac{{\tau_
1} {{{\eta_ 3}}^2}}{2 {{{\tau_ 3}}^3}}-\frac{{\theta_ 1}}{4 {\tau_
3}}+\frac{{\tau_ 1} {\theta_ 3}}{4 {{{\tau_ 3}}^2}} ,}\\& {
{a_{11}}=\frac{{\eta_ 2}}{6 {\tau_ 1}}+\frac{{\tau_ 2} {\eta_ 2}}{6
{\tau_ 1} {\tau_ 3}}-\frac{2 {\eta_ 3}}{3 {\tau_ 1}}-\frac{{{{\tau_
2}}^2} {\eta_ 3}}{3 {\tau_ 1} {{{\tau_ 3}}^2}}-\frac{2 {\tau_ 2}
{\eta_ 3}}{3 {\tau_ 1} {\tau_ 3}}+\frac{2 {\eta_ 4}}{3 {\tau_
1}}+\frac{2 {\tau_ 2} {\eta_ 4}}{3 {\tau_ 1} {\tau_ 3}} ,}\\& {
{a_{12}}=\frac{{\eta_ 2}}{3 {\tau_ 1}}+\frac{{\tau_ 2} {\eta_ 2}}{3
{\tau_ 1} {\tau_ 3}}-\frac{{\eta_ 3}}{3 {\tau_ 1}}-\frac{{\tau_ 2}
{\eta_ 3}}{2 {{{\tau_ 3}}^2}}-\frac{{{{\tau_ 2}}^2} {\eta_ 3}}{6
{\tau_ 1} {{{\tau_ 3}}^2}}-\frac{{\eta_ 3}}{2 {\tau_
3}}-\frac{{\tau_ 2} {\eta_ 3}}{3 {\tau_ 1} {\tau_ 3}}+\frac{{\eta_
4}}{3 {\tau_ 1}}+\frac{{\eta_ 4}}{2 {\tau_ 3}}+\frac{{\tau_ 2}
{\eta_ 4}}{3 {\tau_ 1} {\tau_ 3}} ,}\\& { {a_{13}}=\frac{{\eta_
1}}{6 {\tau_ 1}}+\frac{{\tau_ 2} {\eta_ 1}}{6 {\tau_ 1} {\tau_
3}}-\frac{{\tau_ 1} {\eta_ 3}}{6 {{{\tau_ 3}}^2}}-\frac{{\tau_ 2}
{\eta_ 3}}{6 {{{\tau_ 3}}^2}}-\frac{{\eta_ 3}}{6 {\tau_
3}}+\frac{{\eta_ 4}}{6 {\tau_ 3}} ,}\\& { {a_{14}}=-\frac{{\tau_ 2}
{\eta_ 1}}{3 {{{\tau_ 1}}^2}}-\frac{{{{\tau_ 2}}^2} {\eta_ 1}}{6
{{{\tau_ 1}}^2} {\tau_ 3}}+\frac{{\eta_ 2}}{3 {\tau_
1}}+\frac{{\tau_ 2} {\eta_ 2}}{3 {\tau_ 1} {\tau_ 3}}-\frac{{\eta_
3}}{3 {\tau_ 1}}-\frac{{{{\tau_ 2}}^2} {\eta_ 3}}{6 {\tau_ 1}
{{{\tau_ 3}}^2}}-\frac{{\tau_ 2} {\eta_ 3}}{3 {\tau_ 1} {\tau_
3}}+\frac{{\eta_ 4}}{3 {\tau_ 1}}+\frac{{\tau_ 2} {\eta_ 4}}{3
{\tau_ 1} {\tau_ 3}} ,}\\& { {a_{15}}=\frac{{\eta_ 2}}{3 {\tau_
1}}-\frac{{\tau_ 2} {\eta_ 2}}{3 {{{\tau_ 1}}^2}}+\frac{{\tau_ 2}
{\eta_ 2}}{3 {\tau_ 1} {\tau_ 3}}-\frac{{{{\tau_ 2}}^2} {\eta_ 2}}{6
{{{\tau_ 1}}^2} {\tau_ 3}}-\frac{{\eta_ 3}}{3 {\tau_
1}}-\frac{{{{\tau_ 2}}^2} {\eta_ 3}}{6 {\tau_ 1} {{{\tau_
3}}^2}}-\frac{{\tau_ 2} {\eta_ 3}}{3 {\tau_ 1} {\tau_
3}}+\frac{{\eta_ 4}}{3 {\tau_ 1}}+\frac{{\tau_ 2} {\eta_ 4}}{3
{\tau_ 1} {\tau_ 3}} ,}\\& { {a_{16}}=-\frac{{\eta_ 1}}{6 {\tau_
1}}-\frac{{\tau_ 2} {\eta_ 1}}{6 {\tau_ 1} {\tau_ 3}}+\frac{{\eta_
2}}{6 {\tau_ 1}}+\frac{{\eta_ 2}}{6 {\tau_ 3}}+\frac{{\tau_ 2}
{\eta_ 2}}{6 {\tau_ 1} {\tau_ 3}}-\frac{{\tau_ 2} {\eta_ 3}}{6
{{{\tau_ 3}}^2}}-\frac{{\eta_ 3}}{6 {\tau_ 3}}+\frac{{\eta_ 4}}{6
{\tau_ 3}} ,}\\& { {a_{17}}=\frac{{\eta_ 1}}{6 {\tau_
1}}+\frac{{\tau_ 2} {\eta_ 1}}{6 {\tau_ 1} {\tau_ 3}}-\frac{{\eta_
2}}{6 {\tau_ 1}}+\frac{{\eta_ 2}}{6 {\tau_ 3}}-\frac{{\tau_ 2}
{\eta_ 2}}{6 {\tau_ 1} {\tau_ 3}}-\frac{{\tau_ 2} {\eta_ 3}}{6
{{{\tau_ 3}}^2}}-\frac{{\eta_ 3}}{6 {\tau_ 3}}+\frac{{\eta_ 4}}{6
{\tau_ 3}} ,}
\end{aligned}
$$
$$
\begin{aligned}
 {a_{18}}=&\frac{{\tau_ 2} {\eta_ 1}}{12 {{{\tau_
1}}^2}}+\frac{{{{\tau_ 2}}^3} {\eta_ 1}}{24 {{{\tau_ 1}}^2} {{{\tau_
3}}^2}}+\frac{{{{\tau_ 2}}^2} {\eta_ 1}}{8 {{{\tau_ 1}}^2} {\tau_
3}}-\frac{{\eta_ 2}}{6 {\tau_ 1}}+\frac{{\tau_ 1} {\eta_ 2}}{8
{{{\tau_ 3}}^2}}+\frac{{\tau_ 2} {\eta_ 2}}{24 {{{\tau_
3}}^2}}-\frac{5 {{{\tau_ 2}}^2} {\eta_ 2}}{24 {\tau_ 1} {{{\tau_
3}}^2}}+\frac{{\eta_ 2}}{24 {\tau_ 3}}\\&-\frac{5 {\tau_ 2} {\eta_
2}}{12 {\tau_ 1} {\tau_ 3}}+\frac{5 {\eta_ 3}}{12 {\tau_
1}}-\frac{{\tau_ 1} {\tau_ 2} {\eta_ 3}}{8 {{{\tau_ 3}}^3}}-\frac{5
{{{\tau_ 2}}^2} {\eta_ 3}}{24 {{{\tau_ 3}}^3}}+\frac{{{{\tau_ 2}}^3}
{\eta_ 3}}{3 {\tau_ 1} {{{\tau_ 3}}^3}}-\frac{{\tau_ 1} {\eta_ 3}}{8
{{{\tau_ 3}}^2}}-\frac{5 {\tau_ 2} {\eta_ 3}}{12 {{{\tau_ 3}}^2}}+
\frac{{{{\tau_ 2}}^2} {\eta_ 3}}{{\tau_ 1} {{{\tau_
3}}^2}}\\&-\frac{7 {\eta_ 3}}{24 {\tau_ 3}}+\frac{13 {\tau_ 2}
{\eta_ 3}}{12 {\tau_ 1} {\tau_ 3}}-\frac{5 {\eta_ 4}}{12 {\tau_
1}}+\frac{{\tau_ 1} {\eta_ 4}}{8 {{{\tau_ 3}}^2}}+\frac{7 {\tau_ 2}
{\eta_ 4}}{24 {{{\tau_ 3}}^2}}-\frac{13 {{{\tau_ 2}}^2} {\eta_
4}}{24 {\tau_ 1} {{{\tau_ 3}}^2}}+\frac{7 {\eta_ 4}}{24 {\tau_
3}}-\frac{13 {\tau_ 2} {\eta_ 4}}{12 {\tau_ 1} {\tau_ 3}},
\\  {a_{19}}=&\frac{{\tau_ 2} {\eta_
1}}{24 {{{\tau_ 3}}^2}}+\frac{{\eta_ 1}}{24 {\tau_ 3}}-\frac{{\eta_
2}}{6 {\tau_ 1}}+\frac{{\tau_ 2} {\eta_ 2}}{12 {{{\tau_
1}}^2}}+\frac{{\tau_ 1} {\eta_ 2}}{8 {{{\tau_ 3}}^2}}-\frac{5
{{{\tau_ 2}}^2} {\eta_ 2}}{24 {\tau_ 1} {{{\tau_
3}}^2}}+\frac{{{{\tau_ 2}}^3} {\eta_ 2}}{24 {{{\tau_ 1}}^2} {{{\tau_
3}}^2}}-\frac{5 {\tau_ 2} {\eta_ 2}}{12 {\tau_ 1} {\tau_
3}}\\&+\frac{{{{\tau_ 2}}^2} {\eta_ 2}}{8 {{{\tau_ 1}}^2} {\tau_
3}}+\frac{{\eta_ 3}}{6 {\tau_ 1}}-\frac{{{{\tau_ 1}}^2} {\eta_
3}}{24 {{{\tau_ 3}}^3}}-\frac{7 {\tau_ 1} {\tau_ 2} {\eta_ 3}}{24
{{{\tau_ 3}}^3}}+\frac{{{{\tau_ 2}}^2} {\eta_ 3}}{4 {{{\tau_
3}}^3}}+\frac{{{{\tau_ 2}}^3} {\eta_ 3}}{12 {\tau_ 1} {{{\tau_
3}}^3}}-\frac{7 {\tau_ 1} {\eta_ 3}}{24 {{{\tau_ 3}}^2}}+
\frac{{\tau_ 2} {\eta_ 3}}{2 {{{\tau_ 3}}^2}}\\&+\frac{{{{\tau_
2}}^2} {\eta_ 3}}{4 {\tau_ 1} {{{\tau_ 3}}^2}}+\frac{{\eta_ 3}}{8
{\tau_ 3}}+\frac{{\tau_ 2} {\eta_ 3}}{3 {\tau_ 1} {\tau_
3}}-\frac{{\eta_ 4}}{6 {\tau_ 1}}+\frac{{\tau_ 1} {\eta_ 4}}{6
{{{\tau_ 3}}^2}}-\frac{{\tau_ 2} {\eta_ 4}}{8 {{{\tau_
3}}^2}}-\frac{{{{\tau_ 2}}^2} {\eta_ 4}}{6 {\tau_ 1} {{{\tau_
3}}^2}}-\frac{{\eta_ 4}}{8 {\tau_ 3}}-\frac{{\tau_ 2} {\eta_ 4}}{3
{\tau_ 1} {\tau_ 3}}, \\
 {a_{20}}=&\frac{{\eta_ 1}}{24 {\tau_
1}}+\frac{{{{\tau_ 2}}^2} {\eta_ 1}}{24 {\tau_ 1} {{{\tau_
3}}^2}}+\frac{{\tau_ 2} {\eta_ 1}}{12 {\tau_ 1} {\tau_
3}}-\frac{{\eta_ 2}}{8 {\tau_ 1}}+\frac{{\tau_ 1} {\eta_ 2}}{6
{{{\tau_ 3}}^2}}-\frac{{\tau_ 2} {\eta_ 2}}{12 {{{\tau_
3}}^2}}-\frac{{{{\tau_ 2}}^2} {\eta_ 2}}{8 {\tau_ 1} {{{\tau_
3}}^2}}-\frac{{\eta_ 2}}{12 {\tau_ 3}}-\frac{{\tau_ 2} {\eta_ 2}}{4
{\tau_ 1} {\tau_ 3}}\\&+\frac{{\eta_ 3}}{12 {\tau_
1}}-\frac{{{{\tau_ 1}}^2} {\eta_ 3}}{12 {{{\tau_
3}}^3}}-\frac{{\tau_ 1} {\tau_ 2} {\eta_ 3}}{4 {{{\tau_
3}}^3}}+\frac{7 {{{\tau_ 2}}^2} {\eta_ 3}}{24 {{{\tau_
3}}^3}}+\frac{{{{\tau_ 2}}^3} {\eta_ 3}}{24 {\tau_ 1} {{{\tau_
3}}^3}}-\frac{{\tau_ 1} {\eta_ 3}}{4 {{{\tau_ 3}}^2}}+\frac{7 {\tau_
2} {\eta_ 3}}{12 {{{\tau_ 3}}^2}}+ \frac{{{{\tau_ 2}}^2} {\eta_
3}}{8 {\tau_ 1} {{{\tau_ 3}}^2}}\\&+\frac{7 {\eta_ 3}}{24 {\tau_
3}}+\frac{{\tau_ 2} {\eta_ 3}}{6 {\tau_ 1} {\tau_ 3}}-\frac{{\eta_
4}}{12 {\tau_ 1}}+\frac{{\tau_ 1} {\eta_ 4}}{4 {{{\tau_
3}}^2}}-\frac{7 {\tau_ 2} {\eta_ 4}}{24 {{{\tau_
3}}^2}}-\frac{{{{\tau_ 2}}^2} {\eta_ 4}}{12 {\tau_ 1} {{{\tau_
3}}^2}}-\frac{7 {\eta_ 4}}{24 {\tau_ 3}}-\frac{{\tau_ 2} {\eta_
4}}{6 {\tau_ 1} {\tau_ 3}}, \\  {a_{21}}=&-\frac{{\eta_ 1}}{8 {\tau_
1}}+\frac{{\tau_ 1} {\eta_ 1}}{6 {{{\tau_ 3}}^2}}-\frac{{{{\tau_
2}}^2} {\eta_ 1}}{8 {\tau_ 1} {{{\tau_ 3}}^2}}-\frac{{\tau_ 2}
{\eta_ 1}}{4 {\tau_ 1} {\tau_ 3}}+\frac{{\eta_ 2}}{24 {\tau_
1}}-\frac{{\tau_ 2} {\eta_ 2}}{12 {{{\tau_ 3}}^2}}+\frac{{{{\tau_
2}}^2} {\eta_ 2}}{24 {\tau_ 1} {{{\tau_ 3}}^2}}-\frac{{\eta_ 2}}{12
{\tau_ 3}}+\frac{{\tau_ 2} {\eta_ 2}}{12 {\tau_ 1} {\tau_
3}}\\&-\frac{{{{\tau_ 1}}^2} {\eta_ 3}}{3 {{{\tau_ 3}}^3}}+\frac{5
{\tau_ 1} {\tau_ 2} {\eta_ 3}}{24 {{{\tau_ 3}}^3}}+\frac{{{{\tau_
2}}^2} {\eta_ 3}}{8 {{{\tau_ 3}}^3}}+\frac{5 {\tau_ 1} {\eta_ 3}}{24
{{{\tau_ 3}}^2}}+\frac{{\tau_ 2} {\eta_ 3}}{4 {{{\tau_
3}}^2}}+\frac{{\eta_ 3}}{8 {\tau_ 3}}-\frac{{\tau_ 2} {\eta_ 4}}{8
{{{\tau_ 3}}^2}}-\frac{{\eta_ 4}}{8 {\tau_ 3}}, \\
 {a_{22}}=&\frac{{\tau_ 2} {\eta_ 1}}{12 {{{\tau_
1}}^2}}+\frac{{{{\tau_ 2}}^3} {\eta_ 1}}{24 {{{\tau_ 1}}^2} {{{\tau_
3}}^2}}+\frac{{{{\tau_ 2}}^2} {\eta_ 1}}{8 {{{\tau_ 1}}^2} {\tau_
3}}-\frac{{\eta_ 2}}{12 {\tau_ 1}}+\frac{{\tau_ 1} {\eta_ 2}}{24
{{{\tau_ 3}}^2}}+\frac{{\tau_ 2} {\eta_ 2}}{24 {{{\tau_
3}}^2}}-\frac{{{{\tau_ 2}}^2} {\eta_ 2}}{8 {\tau_ 1} {{{\tau_
3}}^2}}+\frac{{\eta_ 2}}{24 {\tau_ 3}}\\&-\frac{{\tau_ 2} {\eta_
2}}{4 {\tau_ 1} {\tau_ 3}}+\frac{{\eta_ 3}}{12 {\tau_
1}}-\frac{{\tau_ 1} {\tau_ 2} {\eta_ 3}}{12 {{{\tau_
3}}^3}}+\frac{{{{\tau_ 2}}^3} {\eta_ 3}}{12 {\tau_ 1} {{{\tau_
3}}^3}}-\frac{{\tau_ 1} {\eta_ 3}}{12 {{{\tau_
3}}^2}}+\frac{{{{\tau_ 2}}^2} {\eta_ 3}}{4 {\tau_ 1} {{{\tau_
3}}^2}}+\frac{{\tau_ 2} {\eta_ 3}}{4 {\tau_ 1} {\tau_ 3}}-
\frac{{\eta_ 4}}{12 {\tau_ 1}}\\&+\frac{{\tau_ 1} {\eta_ 4}}{24
{{{\tau_ 3}}^2}}-\frac{{{{\tau_ 2}}^2} {\eta_ 4}}{8 {\tau_ 1}
{{{\tau_ 3}}^2}}-\frac{{\tau_ 2} {\eta_ 4}}{4 {\tau_ 1} {\tau_ 3}},
\\ {a_{23}}=&\frac{{\tau_ 2} {\eta_ 1}}{24 {{{\tau_
3}}^2}}+\frac{{\eta_ 1}}{24 {\tau_ 3}}-\frac{{\eta_ 2}}{12 {\tau_
1}}+\frac{{\tau_ 2} {\eta_ 2}}{12 {{{\tau_ 1}}^2}}+\frac{{\tau_ 1}
{\eta_ 2}}{24 {{{\tau_ 3}}^2}}-\frac{{{{\tau_ 2}}^2} {\eta_ 2}}{8
{\tau_ 1} {{{\tau_ 3}}^2}}+\frac{{{{\tau_ 2}}^3} {\eta_ 2}}{24
{{{\tau_ 1}}^2} {{{\tau_ 3}}^2}}-\frac{{\tau_ 2} {\eta_ 2}}{4 {\tau_
1} {\tau_ 3}}\\&+\frac{{{{\tau_ 2}}^2} {\eta_ 2}}{8 {{{\tau_ 1}}^2}
{\tau_ 3}}+\frac{{\eta_ 3}}{12 {\tau_ 1}}-\frac{{\tau_ 1} {\tau_ 2}
{\eta_ 3}}{12 {{{\tau_ 3}}^3}}+\frac{{{{\tau_ 2}}^3} {\eta_ 3}}{12
{\tau_ 1} {{{\tau_ 3}}^3}}-\frac{{\tau_ 1} {\eta_ 3}}{12 {{{\tau_
3}}^2}}+\frac{{{{\tau_ 2}}^2} {\eta_ 3}}{4 {\tau_ 1} {{{\tau_
3}}^2}}+\frac{{\tau_ 2} {\eta_ 3}}{4 {\tau_ 1} {\tau_ 3}}-
\frac{{\eta_ 4}}{12 {\tau_ 1}}\\&+\frac{{\tau_ 1} {\eta_ 4}}{24
{{{\tau_ 3}}^2}}-\frac{{{{\tau_ 2}}^2} {\eta_ 4}}{8 {\tau_ 1}
{{{\tau_ 3}}^2}}-\frac{{\tau_ 2} {\eta_ 4}}{4 {\tau_ 1} {\tau_ 3}},
\\ {a_{24}}=&\frac{{\eta_ 1}}{24 {\tau_ 1}}+\frac{{{{\tau_ 2}}^2}
{\eta_ 1}}{24 {\tau_ 1} {{{\tau_ 3}}^2}}+\frac{{\tau_ 2} {\eta_
1}}{12 {\tau_ 1} {\tau_ 3}}-\frac{{\eta_ 2}}{24 {\tau_
1}}+\frac{{\tau_ 1} {\eta_ 2}}{12 {{{\tau_ 3}}^2}}-\frac{{\tau_ 2}
{\eta_ 2}}{12 {{{\tau_ 3}}^2}}-\frac{{{{\tau_ 2}}^2} {\eta_ 2}}{24
{\tau_ 1} {{{\tau_ 3}}^2}}-\frac{{\eta_ 2}}{12 {\tau_
3}}\\&-\frac{{\tau_ 2} {\eta_ 2}}{12 {\tau_ 1} {\tau_
3}}-\frac{{{{\tau_ 1}}^2} {\eta_ 3}}{12 {{{\tau_
3}}^3}}+\frac{{{{\tau_ 2}}^2} {\eta_ 3}}{12 {{{\tau_
3}}^3}}+\frac{{\tau_ 2} {\eta_ 3}}{6 {{{\tau_ 3}}^2}}+\frac{{\eta_
3}}{12 {\tau_ 3}}-\frac{{\tau_ 2} {\eta_ 4}}{12 {{{\tau_
3}}^2}}-\frac{{\eta_ 4}}{12 {\tau_ 3}}, \\  {a_{25}}=&-\frac{{\eta_
1}}{24 {\tau_ 1}}+\frac{{\tau_ 1} {\eta_ 1}}{12 {{{\tau_
3}}^2}}-\frac{{{{\tau_ 2}}^2} {\eta_ 1}}{24 {\tau_ 1} {{{\tau_
3}}^2}}-\frac{{\tau_ 2} {\eta_ 1}}{12 {\tau_ 1} {\tau_
3}}+\frac{{\eta_ 2}}{24 {\tau_ 1}}-\frac{{\tau_ 2} {\eta_ 2}}{12
{{{\tau_ 3}}^2}}+\frac{{{{\tau_ 2}}^2} {\eta_ 2}}{24 {\tau_ 1}
{{{\tau_ 3}}^2}}-\frac{{\eta_ 2}}{12 {\tau_ 3}}\\&+\frac{{\tau_ 2}
{\eta_ 2}}{12 {\tau_ 1} {\tau_ 3}}-\frac{{{{\tau_ 1}}^2} {\eta_
3}}{12 {{{\tau_ 3}}^3}}+\frac{{{{\tau_ 2}}^2} {\eta_ 3}}{12 {{{\tau_
3}}^3}}+\frac{{\tau_ 2} {\eta_ 3}}{6 {{{\tau_ 3}}^2}}+\frac{{\eta_
3}}{12 {\tau_ 3}}-\frac{{\tau_ 2} {\eta_ 4}}{12 {{{\tau_
3}}^2}}-\frac{{\eta_ 4}}{12 {\tau_ 3}}.
\end{aligned}
$$

\section{}

The formulas of the constants $b_i$ ($i=1,\cdots,22$) in
(3.29)-(3.31) are given below:
$$
\begin{aligned}
{b_1}=&\frac{-5 {\tau_ 1}^3+{\tau_ 2}^2 (-3 {\eta_ 1}+{\eta_
2})-{\tau_ 1} {\tau_ 2} (3 {\eta_ 1}+2 {\eta_ 2})+{\tau_ 1}^2 (-5
{\tau_ 2}+7 {\eta_ 2})}{6 {\tau_ 1} ({\tau_ 1}-{\tau_ 2})},\\
{b_2}=&\frac{5 {\tau_ 1}^3+2 {\tau_ 2}^2 {\eta_ 1}+{\tau_ 1}^2 (5
{\tau_ 2}-7 {\eta_ 2})+{\tau_ 1} {\tau_ 2} (4 {\eta_ 1}+{\eta_
2})}{6 {\tau_ 1} ({\tau_ 1}-{\tau_ 2})},\\
{b_3}=&\frac{5 {\tau_ 1}^3+{\tau_ 1}^2 (5 {\tau_ 2}+2 {\eta_ 1}-9
{\eta_ 2})+2 {\tau_ 2}^2 {\eta_ 2}+{\tau_ 1} {\tau_ 2} (4 {\eta_
1}+{\eta_ 2})}{6 {\tau_ 1} ({\tau_ 1}-{\tau_ 2})},\\
{b_4}=&\frac{-5 {\tau_ 1}^3+{\tau_ 1} {\tau_ 2} (-6 {\eta_ 1}+{\eta_
2})+{\tau_ 2}^2 (-3 {\eta_ 1}+{\eta_ 2})+{\tau_ 1}^2 (-5 {\tau_ 2}+3
{\eta_ 1}+4 {\eta_ 2})}{6 {\tau_ 1} ({\tau_ 1}-{\tau_
2})},\\
{b_5}=&\frac{{\tau_ 1}^3+{\tau_ 1}^2 ({\tau_ 2}-3 {\eta_ 2})+{\tau_
2}^2 (-{\eta_ 1}+{\eta_ 2})+{\tau_ 1} {\tau_ 2} ({\eta_ 1}+2 {\eta_
2})}{6 {\tau_ 1} ({\tau_ 1}-{\tau_ 2})},\\
{b_6}=&\frac{-{\tau_ 1}^3-3 {\tau_ 1} {\tau_ 2} {\eta_ 2}+{\tau_
2}^2 (-{\eta_ 1}+{\eta_ 2})+{\tau_ 1}^2 (-{\tau_ 2}+{\eta_ 1}+2
{\eta_ 2})}{6 {\tau_ 1} ({\tau_ 1}-{\tau_ 2})},\\
{b_7}=&\frac{{\tau_ 1}^3+{\tau_ 1}^2 ({\tau_ 2}+{\eta_ 1}-4 {\eta_
2})+3 {\tau_ 1} {\tau_ 2} {\eta_ 2}+{\tau_ 2}^2 (-{\eta_ 1}+{\eta_
2})}{6 {\tau_ 1} ({\tau_ 1}-{\tau_ 2})},\\
{b_8}=&\frac{{\tau_ 1}^3+{\tau_ 1}^2 ({\tau_ 2}-3 {\eta_ 1})+{\tau_
2}^2 ({\eta_ 1}-{\eta_ 2})+{\tau_ 1} {\tau_ 2} (2 {\eta_ 1}+{\eta_
2})}{6 {\tau_ 1} ({\tau_ 1}-{\tau_ 2})},\\
{b_9}=&\frac{{\tau_ 1}^2+{\tau_ 1} {\tau_ 2}-{\tau_ 1} {\eta_
2}+{\tau_ 2} {\eta_ 2}}{2 {\tau_ 1}-2 {\tau_ 2}},\\
{b_{10}}=&\frac{{\tau_ 1}^2+{\tau_ 1} {\tau_ 2}-{\tau_ 1} {\eta_
2}+{\tau_ 2} {\eta_ 2}}{{\tau_ 1}-{\tau_ 2}},\\
{b_{11}}=&\frac{{\tau_ 1}^2+{\tau_ 1} {\tau_ 2}-{\tau_ 1} {\eta_
1}+{\tau_ 2} {\eta_ 1}}{2 {\tau_ 1}-2 {\tau_ 2}},\\
{b_{12}}=&\frac{{\tau_ 1}^2+{\tau_ 1} {\tau_ 2}-{\tau_ 1} {\eta_
2}+{\tau_ 2} {\eta_ 2}}{2 {\tau_ 1}-2 {\tau_ 2}},\\
{b_{13}}=&\frac{{\tau_ 1}^2+{\tau_ 1} {\tau_ 2}-{\tau_ 1} {\eta_
2}+{\tau_ 2} {\eta_ 2}}{2 {\tau_ 1}-2 {\tau_ 2}},\\
{b_{14}}=&\frac{{\tau_ 1}^2+{\tau_ 1} {\tau_ 2}-{\tau_ 1} {\eta_
2}+{\tau_ 2} {\eta_ 2}}{2 {\tau_ 1}-2 {\tau_ 2}},\\
{b_{15}}=&\frac{{\tau_ 1}^2+{\tau_ 1} {\tau_ 2}-{\tau_ 1} {\eta_
1}+{\tau_ 2} {\eta_ 1}}{2 {\tau_ 1}-2 {\tau_ 2}},\\
{b_{16}}=&\frac{2 {\tau_ 1}^3+{\tau_ 1} {\tau_ 2} {\eta_ 1}+2 {\tau_
1}^2 ({\tau_ 2}-{\eta_ 2})+{\tau_ 2}^2 (2 {\eta_ 1}-{\eta_ 2})}{6
{\tau_ 1} ({\tau_ 1}-{\tau_ 2})},
\end{aligned}
$$
$$
\begin{aligned}
{b_{17}}=&\frac{4 {\tau_ 1}^3+{\tau_ 1}^2 (4 {\tau_ 2}+{\eta_ 1}-5
{\eta_ 2})+{\tau_ 1} {\tau_ 2} (4 {\eta_ 1}-2 {\eta_ 2})+{\tau_ 2}^2
({\eta_ 1}+{\eta_ 2})}{6 {\tau_ 1} ({\tau_ 1}-{\tau_ 2})},\\
{b_{18}}=&\frac{2 {\tau_ 1}^3+{\tau_ 2}^2 {\eta_ 1}+2 {\tau_ 1}^2
({\tau_ 2}-{\eta_ 2})+{\tau_ 1} {\tau_ 2} (2 {\eta_ 1}-{\eta_ 2})}{6
{\tau_ 1} ({\tau_ 1}-{\tau_ 2})},\\
{b_{19}}=&\frac{{\tau_ 1}^3+{\tau_ 1}^2 ({\tau_ 2}-3 {\eta_
2})+{\tau_ 2}^2 (-{\eta_ 1}+{\eta_ 2})+{\tau_ 1} {\tau_ 2} ({\eta_
1}+2 {\eta_ 2})}{6 {\tau_ 1} ({\tau_ 1}-{\tau_ 2})},\\
{b_{20}}=&\frac{-{\tau_ 1}^3-3 {\tau_ 1} {\tau_ 2} {\eta_ 2}+{\tau_
2}^2 (-{\eta_ 1}+{\eta_ 2})+{\tau_ 1}^2 (-{\tau_ 2}+{\eta_ 1}+2
{\eta_ 2})}{6 {\tau_ 1} ({\tau_ 1}-{\tau_ 2})},\\
{b_{21}}=&\frac{{\tau_ 1}^3+{\tau_ 1}^2 ({\tau_ 2}+{\eta_ 1}-4
{\eta_ 2})+3 {\tau_ 1} {\tau_ 2} {\eta_ 2}+{\tau_ 2}^2 (-{\eta_
1}+{\eta_ 2})}{6 {\tau_ 1} ({\tau_ 1}-{\tau_ 2})},\\
{b_{22}}=&\frac{{\tau_ 1}^3+{\tau_ 1}^2 ({\tau_ 2}-3 {\eta_
1})+{\tau_ 2}^2 ({\eta_ 1}-{\eta_ 2})+{\tau_ 1} {\tau_ 2} (2 {\eta_
1}+{\eta_ 2})}{6 {\tau_ 1} ({\tau_ 1}-{\tau_ 2})}.
\end{aligned}
$$

\section{}

The formulas of the constants $c_i$ ($i=1,\cdots,12$) in (3.38) are given below:
$$
\begin{aligned}
c_1=&-\frac{\tau_1^3+\tau_1 \tau_2 (\tau_2-2 \eta_2)+\tau_1^2 (2
\tau_2-\eta_2)+\tau_2^2 (2
\eta_1+\eta_2)}{6 (\tau_1-\tau_2)^2},\\
c_2=&\frac{1}{6 \tau_1 (\tau_1-\tau_2)^2} (2 \tau_1^4-2 \tau_1^3
(\tau_2+\eta_2)+\tau_2^3 (-2 \eta_1+\eta_2)-\tau_1 \tau_2^2
(\eta_1+3
\eta_2)\\&+\tau_1^2 \tau_2 (-4 \tau_2+\eta_1+6 \eta_2)),\\
c_3=&-\frac{2 \tau_1^3+2 \tau_1 \tau_2
\eta_1+\tau_1^2 (2 \tau_2-3 \eta_2)+\tau_2^2 \eta_2}{3 (\tau_1-\tau_2)^2}, \\
c_4=&\frac{1}{6 \tau_1 (\tau_1-\tau_2)^2} (2 \tau_1^4+\tau_1^3 (-4
\tau_2+\eta_1-\eta_2)-\tau_2^3 (\eta_1+\eta_2)\\&+3 \tau_1^2 \tau_2 (-2
\tau_2+\eta_1+\eta_2)+\tau_1 \tau_2^2 (-7 \eta_1+3 \eta_2)),
\\
{c_5}=&\frac{1}{6 ({\tau_1}-{\tau_2})^2}\left(-3
{\tau_1}^3+{\tau_2}^2 {\eta_1}-{\tau_1}^2 (2 {\tau_2}+{\eta_1}-6
{\eta_2})+{\tau_1} {\tau_2} ({\tau_2}-2 ({\eta_1}+2 {\eta_2}))\right),\\
{c_6}=&\frac{1}{6 {\tau_1} ({\tau_1}-{\tau_2})^2} (-2 {\tau_1}^3
({\tau_2}-{\eta_1})-{\tau_2}^3 {\eta_1}+{\tau_1}^2 {\tau_2} (-2
{\tau_2}-2 {\eta_1}+{\eta_2})\\&+{\tau_1} {\tau_2}^2
(-{\eta_1}+{\eta_2})),\\
{c_7}=&\frac{2 {\tau_1}^2+2 {\tau_1} {\tau_2}+2 {\tau_2}
{\eta_1}-3 {\tau_1} {\eta_2}+{\tau_2} {\eta_2}}{-12
{\tau_1}+12 {\tau_2}},\\
{c_8}=&\frac{{\tau_2} {\eta_1}-{\tau_1} {\eta_2}-{\tau_2}
{\eta_2}}{6 {\tau_1}},\\
{c_9}=&\frac{1}{12 {\tau_1} ({\tau_1}-{\tau_2})}\left(2
{\tau_1}^3+{\tau_1}^2 (2 {\tau_2}-{\eta_1}-4 {\eta_2})+2 {\tau_2}^2
({\eta_1}-{\eta_2})+{\tau_1} {\tau_2} ({\eta_1}+4 {\tau_2})\right),\\
{c_{10}}=&\frac{2 {\tau_1}^2+{\tau_1} (2 {\tau_2}+{\eta_1}-4
{\eta_2})+{\tau_2} ({\eta_1}+2 {\eta_2}) }{12
(-{\tau_1}+{\tau_2})},\\
{c_{11}}=&\frac{2 {\tau_1} {\eta_1}+{\tau_2} {\eta_1}-3 {\tau_1}
{\eta_2}-{\tau_2} {\eta_2}}{6 {\tau_1}},\\
{c_{12}}=&\frac{1}{12 {\tau_1} ({\tau_1}-{\tau_2})}\left(2
{\tau_1}^3+{\tau_1}^2 (2 {\tau_2}-4 {\eta_1}-{\eta_2})+2 {\tau_2}^2
({\eta_1}-{\eta_2})+{\tau_1} {\tau_2} (4 {\eta_1}+{\eta_2})\right).
\end{aligned}
$$



\end{document}